\documentclass[aps,prx,twocolumn,superscriptaddress, nofootinbib,  longbibliography]{revtex4-2}
\usepackage{graphicx}
\usepackage{dcolumn} 
\usepackage{bm}   
\usepackage{amssymb} 
\usepackage{amsfonts, amsmath, amssymb, mathtools}
\usepackage{amsmath}

\usepackage{lipsum}
\usepackage[normalem]{ulem}
\usepackage{commath}
\usepackage{xcolor}
\usepackage{comment}
\usepackage[toc,page]{appendix}
\usepackage{booktabs}
\usepackage[normalem]{ulem}
\usepackage{lineno}
\usepackage[binary-units=true]{siunitx}
\usepackage{hyperref}
\usepackage{cleveref}
\RequirePackage{textcase}

\begin{document}

\raggedbottom

\title{A direct controlled-phase gate between microwave photons}

\author{Adrian Copetudo}
\affiliation{Centre for Quantum Technologies, National University of Singapore, Singapore}
\author{Amon M. Kasper}
\affiliation{Centre for Quantum Technologies, National University of Singapore, Singapore}
\author{Tanjung Krisnanda}
\affiliation{Centre for Quantum Technologies, National University of Singapore, Singapore}
\author{Gregoire Veyrac}
\affiliation{Centre for Quantum Technologies, National University of Singapore, Singapore}
\author{Shushen Qin}
\affiliation{Centre for Quantum Technologies, National University of Singapore, Singapore}
\author{Hui Khoon Ng}
\affiliation{Department of Physics, National University of Singapore, Singapore}
\affiliation{Centre for Quantum Technologies, National University of Singapore, Singapore}
\author{Yvonne Y. Gao}
\email[Corresponding author: ]{yvonne.gao@nus.edu.sg}
\affiliation{Centre for Quantum Technologies, National University of Singapore, Singapore}
\affiliation{Department of Physics, National University of Singapore, Singapore}
\date{\today}

\begin{abstract}
The rich dynamics and large Hilbert space of quantum harmonic oscillators make them natural candidates for hardware-efficient and error-correctable quantum information processing. However, implementing direct entangling operations between oscillators remains an outstanding challenge. Existing strategies typically rely on parametrically activating interactions that populate the excited states of a nonlinear element, which introduces additional dissipation channels and potential leakage from the encoded manifold. Here, we engineer a Raman-assisted cross-Kerr interaction between microwave photons hosted in two superconducting cavities. Crucially, this dynamics does not excite the mediating nonlinear coupler, thereby suppressing coupler induced decoherence and leakage out of the bosonic code space. We use this direct nonlinear coupling to implement a controlled-phase gate within the single- and two-photon subspaces of two oscillators, deterministically generating entanglement between them. Finally, we use these engineered dynamics to implement a photon-number parity check on a storage cavity via purely bosonic interactions with an ancillary cavity, demonstrating an enhancement in the storage lifetime. Our work provides a promising pathway toward engineering robust operations that act entirely within a protected bosonic code space and realizing fault-tolerant quantum information processing with bosonic elements.
\end{abstract}

\maketitle

Quantum harmonic oscillators represent one of the most attractive platforms for quantum information processing. Whether manifested as stationary or propagating electromagnetic modes, or as phononic excitations of mechanical motion, their naturally weak environmental coupling and inherently large Hilbert space enable hardware-efficient encoding of quantum information. These intrinsic advantages have driven landmark achievements across quantum communication~\cite{furusawa1998unconditional, liao2017satellite, konno2023propagating}, metrology~\cite{tse2019quantum, jia2024squeezing, pan2025realization}, and quantum error correction (QEC) beyond the break-even point in multiple physical platforms~\cite{sivak2023real, shirol2025passive, larsen2025integrated, konno2024logical}, establishing bosonic systems as efficient carriers of quantum information and essential building blocks for robust quantum computation.

However, it is still an outstanding challenge to create direct nonlinear coupling between bosonic modes~\cite{chang2014quantum} - a key ingredient for deterministic entanglement and universal quantum computation. Two harmonic oscillators in close proximity to one another naturally hybridize into linear normal modes that remain non-interacting. Consequently, deterministic entanglement is typically implemented using parametrically driven dynamics mediated by excitations in a nonlinear ancillary element~\cite{chuang1995simple, paternostro2003generation, ding2017cross, chen2021quantum}. For instance, in existing bosonic circuit quantum electrodynamics (cQED) demonstrations, each harmonic mode is entangled with the excited states of a nonlinear coupler sequentially to engineer an effective interaction between the oscillators~\cite{rosenblum2018cnot, gao2019entanglement, xu2020demonstration, reuer2022realization}. Although successful, such schemes necessarily lead to leakage outside the protected bosonic code space and introduce additional decoherence channels for the oscillators, limiting the gate fidelity and fault-tolerance~\cite{gao2018programmable, zhang2019engineering, goldblatt2024recovering}.

In this work, we implement a direct Raman-assisted cross-Kerr interaction between microwave photons hosted in two superconducting cavities without populating the mediating transmon coupler. This engineered coupling, which can be activated on-demand, is two orders of magnitude stronger than the residual always-on cross-Kerr. We use this nonlinear dynamics to perform a controlled-phase (CPHASE) gate in the single- and two-photon subspace with an average imperfection per gate below $5\%$, 
limited mainly by oscillator decoherence. 
Importantly, our technique generates entanglement directly between the two bosonic fields without populating the excited states of the nonlinear coupler, preserving the code space throughout the operation. Further, we leverage this cross-Kerr coupling to perform a bosonic parity check on a storage cavity in a biased-erasure encoding~\cite{sahay2023high, mai2026biased}. Heralded by the state of an ancillary cavity, our method enhances the storage lifetime without subjecting it to interactions with auxiliary nonlinear elements. This direct photon–photon interaction provides an ingredient for bosonic operations that suppress coupler-induced decoherence and preserve the encoded subspace, making our method inherently compatible with a wide range of existing bosonic QEC methods~\cite{cai2021bosonic}. Overall, our engineered cross-Kerr coupling between microwave photons offers a promising path to explore nonlinear quantum optical effects and realize fault-tolerant information processing using error-correctable bosonic modes~\cite{susan}.

\begin{figure}[!t]
\centering
\includegraphics{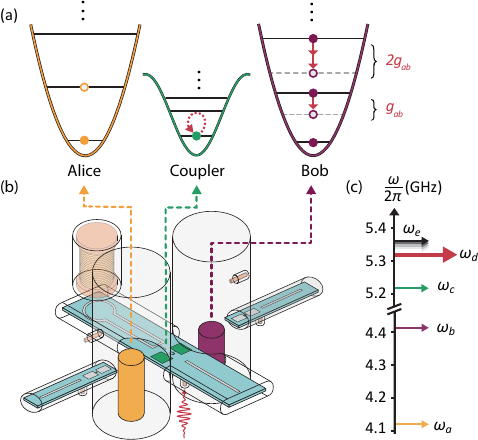}
\caption{
    \textbf{Driven cross-Kerr protocol and experimental device.} 
    (a) Conceptual illustration of the engineered direct cross-Kerr, $g_{ab}$, between two harmonic oscillators mediated by a nonlinear coupler. A photon in Alice shifts the energy levels of Bob without physically populating the coupler.
    (b) Schematic of the experimental device, consisting of two 3D superconducting cavities, Alice and Bob, each coupled to its own auxiliary transmon used for state preparation and tomography. A central flux-tunable transmon acts as the coupler and is biased by a superconducting coil via a pick-up loop transformer.
    (c) Frequency arrangement used in this work. A strong pump is placed at $\omega_d$ with a detuning $\Delta/2\pi$ of a few MHz from the resonance exchange $|1\rangle|0\rangle|g\rangle \leftrightarrow |0\rangle|1\rangle|e\rangle$ occurring at $\omega_e/2\pi\, \mathord{=}\, 5.34\,$GHz. Higher photon resonances are shifted down in frequency by multiples of $\chi_a$ and $\chi_b$. }
    \label{fig:fig1_device}
\end{figure}

Generally, the cross-Kerr interaction between two oscillators, Alice and Bob, takes the form
\begin{equation}
\hat H_{\chi}/\hbar\,\mathord{=} \,g_{ab}\, \hat a^\dagger \hat a \,\hat b^\dagger \hat b, 
\label{Eq: cross_kerr} 
\end{equation} 
where $\hat a$ and $\hat b$ are the annihilation operators of the two cavity modes, and $g_{ab}$ denotes the cross-Kerr coupling strength. Intuitively, this is a direct nonlinear coupling between two bosonic elements where the resonance frequency of one depends on the photon population of the other, as illustrated in Fig.\,\ref{fig:fig1_device}(a).  

In the context of bosonic cQED systems, the oscillators are typically realized using high-Q aluminum superconducting stub cavities, as shown in Fig.\,\ref{fig:fig1_device}(b). Here, Alice and Bob, with frequencies $\omega_a/2\pi\, \mathord{=}\, 4.12\,$GHz and $\omega_b/2\pi \,\mathord{=}\, 4.41\,$GHz, are dispersively coupled to a common nonlinear coupler in the form of an asymmetric Superconducting Quantum Interference Device (SQUID), with strength $\chi_a$ and $\chi_b$, respectively. The coupler frequency, $\omega_c/2\pi$, can be tuned from $4.3\,$GHz to $5.7\,$GHz by a DC magnetic flux applied via a superconducting transformer loop~\cite{zimmerman1971sensitivity, chapman2023high} from a distant coil. This architecture effectively preserves cavity coherences, with single photon lifetimes exceeding 700\,$\mu$s in both Alice and Bob. In such systems, $H_{\chi}$ arises natively due to the mixing of the two oscillator bosonic fields mediated by the nonlinear coupler, whose dynamics is akin to a third-order nonlinear medium. However, this always-on interaction is generally undesirable and carefully minimized. 

To achieve a strong on-demand cross-Kerr without populating the nonlinear coupler, we off-resonantly drive a coherent exchange interaction between the states $|1\rangle_a|0\rangle_b|g\rangle_c$ and $|0\rangle_a|1\rangle_b|e\rangle_c$ to activate the dynamics given by
\begin{equation}
    \hat H_d/\hbar\,\mathord{=} \,g_1 \left( e^{i\,\Delta\,t } \,\hat a^\dagger \hat b \, |g\rangle\langle e| + \text{h.c.} \right), 
    \label{Eq: coherent exchange}
\end{equation}
where $|g\rangle \left(|e\rangle\right)$ denote the ground (excited) state of the coupler, $g_1$ the exchange rate, and $\Delta$ the detuning between the drive frequency, $\omega_d$, and the resonant exchange frequency, $\omega_e \,\mathord{=}\, \omega_c + \omega_b - \omega_a - \chi_b$. 

In the Raman regime, where the drive is far detuned from resonance by $\Delta\, \mathord{\gg} \,g_1$, the interaction terms of Eq.\,\ref{Eq: coherent exchange} oscillate rapidly and are eliminated to first order under the rotating-wave approximation. The dynamics are thus governed by second-order processes in which the state $|1\rangle|0\rangle|g\rangle$ only virtually transitions through $|0\rangle|1\rangle|e\rangle$. We treat this off-resonant interaction perturbatively via a Schrieffer–Wolff transformation~\cite{schrieffer1966relation, sm} to derive the effective cross-Kerr Hamiltonian,
\begin{equation}
    \hat H_\text{eff}/\hbar\,\mathord{=} \,g_{ab} \hat a^\dagger \hat a \,\hat b^\dagger \hat b\,|g\rangle\langle g|,
    \label{Eq: engineered cross_kerr}
\end{equation}
where the strength of the engineered coupling is given by
\begin{equation}
    g_{ab} \approx \frac{g_1^2}{\Delta} \frac{\Delta-\chi_{b}}{\Delta+\chi_{b}}.
    \label{Eq: gcz_SWT}
\end{equation}

\begin{figure}[!t]
    \centering
    \includegraphics{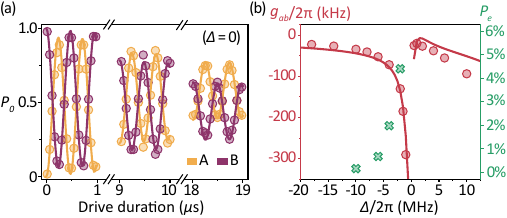}
    \caption{
    \textbf{Calibration of the cross-Kerr protocol parameters.}
    (a) Coherent exchange of population between $|1\rangle|0\rangle|g\rangle$ and $|0\rangle|1\rangle|e\rangle$ modes when $\Delta\,\mathord{=}\,0$ obtained by monitoring the vacuum population of Alice (yellow) and Bob (purple). Oscillation frequency corresponds to $g_1/2\pi \,\mathord{=}\, 1.024\,\mathord{\pm}\,0.004$\,MHz. The data (circles) agrees well with master equation simulations (lines) with the independently measured cavity and coupler decoherence. 
    (b) Strength of engineered cross-Kerr (red), extracted from cavity Ramsey experiments, and residual coupler excitations (green) as a function of $\Delta$, for a particular $g_1/2\pi\,\mathord{\approx}\,0.31\,$MHz. Solid line is a Floquet simulation using experimentally extracted Hamiltonian parameters~\cite{sm}.}
    \label{fig:fig2_exchange}
\end{figure}

We experimentally calibrate $g_1$ by loading a photon in Alice and driving the coupler at a frequency $\omega_e$. This activates the coherence exchange that leads to oscillations of the cavity populations as a function of the drive duration. An exemplary measurement is shown in Fig.\,\ref{fig:fig2_exchange}(a). This time dynamics is consistent with a full master equation simulation with system decoherence taken into account~\cite{sm}. The exchange rate extracted from the period of this oscillation, $g_1/2\pi \,\mathord{=}\, 1.024\,\mathord{\pm}\,0.004\,$MHz, is in good agreement with the analytically predicted value of $g_1/{2\pi} \,\mathord{\approx}\, \sqrt{\chi_{a}\,\chi_{b}}\,\xi /{2\pi}\,\mathord{\approx}\, 1.01\,$MHz, where we used the independently measured dispersive shifts $\chi_{a}/{2\pi}\,\mathord{\approx}\, 2.8\,$MHz and $\chi_{b}/{2\pi}\,\mathord{\approx}\, 3.0\,$MHz, and an effective drive strength $|\xi|\,\mathord{\approx}\, 0.35\,$, obtained from coupler AC Stark shift measurements~\cite{sm}. 

We leverage the flux-tunability of the coupler to optimize the frequency arrangements and suppress the native always-on cross-Kerr interaction, $K_{ab}$, between Alice and Bob, while maintaining a sizable $g_1/2\pi$ $\sim$\,1\,MHz~\cite{sm}. This ensures that the engineered cross-Kerr dynamics has a high on-off ratio, an important metric for high-quality quantum gates. With these considerations in mind, we bias the coupler at a frequency of $\omega_c/2\pi \,\mathord{=}\, 5.22\,$GHz for the rest of this work, where we can realize an exchange rate up to $g_1/2\pi\, \mathord{=}\, 0.8\,$MHz with a weak static cross-Kerr of $K_{ab}/2\pi\,\mathord{\approx}\, 0.3\,$kHz. 

Next, we optimize the drive detuning $\Delta$ by considering the trade-off between the residual coupler excitations and the gate speed. Intuitively, a smaller $\Delta$ allows for a stronger cross-Kerr and thus, faster entangling operation. However, when $\Delta \,\mathord{\sim}\, g_1$, the terms in $\hat H_d$ have a sizable contribution in the first order RWA, resulting in real excitations of the coupler mode. The upper bound of the coupler excitation, $P_e^{\text{max}}$, is given by 
\begin{equation}
    P_e^\text{max}\,\mathord{=} \,\frac{g_1^2}{g_1^2+\Delta^2}.
    \label{Eq: Pe}
\end{equation}

Experimentally, we calibrate $g_{ab}$ at different detunings $\Delta$ with a Ramsey sequence that compares the phase accumulated by Alice when Bob is either in $|0\rangle$ or $|1\rangle$. The experimentally extracted cross-Kerr strength is shown in Fig.\,\ref{fig:fig2_exchange}(b) at both positive and negative $\Delta$. We observe that, in agreement with Eq.\,\ref{Eq: gcz_SWT}, $|g_{ab}|/2\pi$ increases from $\sim$\,10\,kHz to $\sim$\,300\,kHz as $\Delta/2\pi$ is tuned from $-20\,$MHz to $-2\,$MHz. Interestingly, the asymmetry of the engineered cross-Kerr for $\Delta\,\mathord{>}\,0$ arises from the interplay between the drive detuning and the dispersive shift between the coupler and Bob~\cite{sm}. We measure the corresponding average residual populations of the coupler after a $\sim$\,5$\,\mu$s drive at different detunings, as shown by the green markers in Fig.\,\ref{fig:fig2_exchange}(b). Based on this, we choose an operating point of $\Delta/2\pi\,\mathord{=}\,\mathord{-}6\,$MHz, which yields $g_{ab}/2\pi\,\mathord{\approx}\,90\,$kHz. This configuration guarantees a large on-off ratio of $g_{ab}/K_{ab}\gtrsim 250$ while ensuring minimal coupler excitations below $P_e\,\mathord{<}\,1\%$. 

\begin{figure*}[!tbh]
    \centering
    \includegraphics{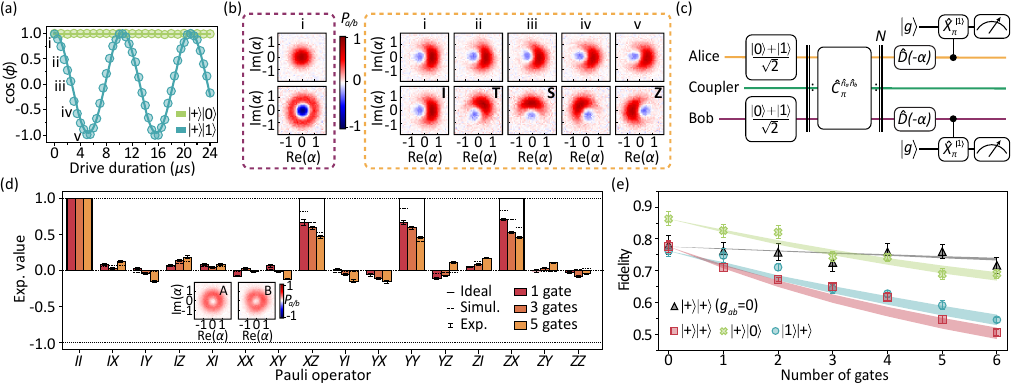}
    \caption{
        \textbf{CPHASE gate in the single-photon manifold.}
        (a) Phase accumulated by Alice, initialized in $|+\rangle \,\mathord{=}\,(|0\rangle+|1\rangle)/{\sqrt{2}}$, as a function of the evolution time under $H_{\text{eff}}$, with Bob in state $|0\rangle$ (green) or $|1\rangle$ (turquoise). A full CZ gate is implemented in 5.244\,$\mathord{\pm}$\,0.008\,$\mu$s.
        (b) Sampled Wigner functions of Alice (yellow) and Bob (purple) at specific times corresponding to a controlled-S, T and Z gates.
        (c) Protocol used to obtain a maximally-entangled state, where cavities are prepared in $|\text{+}\rangle|\text{+}\rangle$. The CZ gates, $\hat C_\pi^{\hat n_a \hat n_b}$, are applied $N$ times, followed by state tomography, where the oscillator populations in $|1\rangle$ are sampled for a set of optimized displacements, $\alpha$, in phase space~\cite{sm}.
        (d) Pauli basis representation of the reconstructed two-cavity state after applying $N\,\mathord{=}\,1,3,5$ gates to $|\text{+}\rangle|\text{+}\rangle$. Inset shows the Wigner functions of the statistical mixture in the oscillators when they are measured independently after one CZ gate.
        (e) Repeated application of $\hat C_\pi^{\hat n_a \hat n_b}$ across a set of initial states. Dots are raw experimental data without any scaling. The $N\,\mathord{=}\,0$ data points indicate that fidelity is limited by state preparation and measurement fidelity. The control experiment, where $|\text{+}\rangle|\text{+}\rangle$ is prepared and measured after an idle time (black triangles), indicates an infidelity of 0.7$\,\mathord{\pm}\,0.3\%$ per gate duration. When the gate is activated, $|\text{+}\rangle|\text{+}\rangle$ (red squares) introduces an infidelity of 4.2$\,\mathord{\pm}\,0.3\%$ per gate. Similarly, states $|\text{+}\rangle|0\rangle$ (green crosses) and $|1\rangle|\text{+}\rangle$ (blue circles) suffer an infidelity of 3.2$\,\mathord{\pm}\,0.4\%$ and 3.9$\,\mathord{\pm}\,0.2\%$ per gate, respectively. Shaded regions represent master equation simulations of the dynamics given by Eq.\,\ref{Eq: engineered cross_kerr}. Spread corresponds to $\mathord{\pm}\,10\%$ fluctuation of cavity coherences, as observed experimentally. 
    }
    \label{fig:fig3_01}
\end{figure*}

This engineered cross-Kerr coupling functions as a direct controlled-phase (CPHASE) operation for bosonic states in the 0/n subspace, which forms a biased-erasure encoding with logical codewords of the form $\{|0\rangle, |n\rangle \}$ (with $n \ge 1$). Concretely, the time evolution under Eq.\,\ref{Eq: engineered cross_kerr} for a duration $T$ yields a non-zero phase only to the state
\begin{equation} 
|1\rangle_L|1\rangle_L\,\mathord{=} \,|n_a\rangle|n_b\rangle \rightarrow e^{-ig_{ab}n_an_bT} |n_a\rangle|n_b\rangle \,\mathord{=}\, e^{i\phi}|1\rangle_L|1\rangle_L, 
\end{equation} 
where $\phi \,\mathord{=}\,\mathord{-}g_{ab}n_an_bT$ is a continuous controlled-phase that can be tuned by changing the interaction time. More broadly, for any rotationally-symmetric bosonic code~\cite{grimsmo2020quantum}, this interaction can impart a controlled-$Z$ gate by tuning the interaction strength and time such that $g_{ab}T \,\mathord{=}\, \frac{\pi}{NM}$, where $N$ and $M$ are the rotational symmetries of each cavity encoding. This condition ensures that only the logical $|1\rangle_L|1\rangle_L$ component acquires a full $\pi$ phase, while all other logical basis states remain unchanged as they acquire phases that are integer multiples of $2\pi$.

As an example, we characterize the CPHASE gate within the 0/1 photon subspace, often used to implement dual-rail qubits~\cite{teoh2023dual, chou2024superconducting}. We prepare Alice and Bob in different basis states and activate the cross-Kerr interaction for a variable duration. The state in Alice, initialized as $|+\rangle \,\mathord{=}\, (|0\rangle + |1\rangle)/\sqrt{2}$, remains unchanged under this dynamics when Bob is in $|0\rangle$, but acquires a continuous phase when Bob is in $|1\rangle$. The phase accumulated on Alice, extracted from the reconstructed density matrices~\cite{sm} after each evolution time, is shown in Fig.\,\ref{fig:fig3_01}(a). A cosine fit (solid line) indicates a controlled-Z (CZ) gate time of approximately 5.2\,$\mu$s. Moreover, by simply varying the gate duration, we realize an effective controlled-T ($\phi \,\mathord{=}\, \pi/4$) and controlled-S ($\phi \,\mathord{=} \,\pi/2$) gates, at times $T\,\mathord{=} \,1.3\,\mu$s and $2.6\,\mu$s, respectively. The Wigner functions measured at a few selected times with Bob initialized in either $|0\rangle$ or $|1\rangle$, as shown in Fig.\,\ref{fig:fig3_01}(b), exhibit phase-space rotations corresponding to a CPHASE operation. 

Next, we apply the calibrated CZ gate to Alice and Bob, both prepared in the superposition state $|\text{+}\rangle|\text{+}\rangle$, and probe the resulting entanglement between the microwave photons, as illustrated in Fig.\,\ref{fig:fig3_01}(c). The reconstructed state, which is a Bell pair of the form $\frac{1}{2}\left( |0\rangle|0\rangle + |0\rangle|1\rangle + |1\rangle|0\rangle - |1\rangle|1\rangle\right)$, is shown in the Pauli-operator representation in Fig.\,\ref{fig:fig3_01}(d). The solid bars show the experimentally reconstructed values, which generally follow the ideal expectation values (solid line) and are consistent with the outcomes of master-equation simulations (dotted lines). The inset displays the measured single-cavity Wigner functions of the maximally entangled state after one CZ gate. Each reduced state exhibits no visible phase structure and closely resembles a maximally-mixed state, with the expected loss of single-mode coherence upon tracing out the other subsystem. In addition, we also include the resulting states after 3 and 5 CZ operations. Notably, the IZ and ZI components grow steadily with the number of gates, indicating that the dominant form of imperfection in our gate is photon loss from Alice and Bob. The other spurious components are consistent with imperfections in the initial state preparation and small miscalibrations of the gate duration. 

Finally, we analyze the gate quality on different basis states in the 0/1 encoding by tracking the state fidelities after applying multiple CZ operations. From a linear fit to the fidelity decay, we extract an average infidelity per gate of $3.2\,\mathord{\pm}\,0.4$\%, $3.9\,\mathord{\pm}\,0.2$\% and $4.3\,\mathord{\pm}\,0.3$\%, when the control is in $|0\rangle$, $|1\rangle$ or $|+\rangle$, respectively. The infidelity is dominated by cavity decoherence during the gate, which is pronounced due to drive-induced decay~\cite{carde2025flux, boissonneault2009dispersive, dai2026characterization} and dressed dephasing~\cite{boissonneault2008nonlinear, gambetta2006qubit}. Our results are in good agreement with the outcome of full master equation simulations (shaded regions in Fig.\,\ref{fig:fig3_01}(e)), where we simulate the evolution of the experimentally reconstructed initial state under the cross-Kerr dynamics that include driven cavity decoherences. The width of the shaded bands corresponds to a $\mathord{\pm} \,10\%$ variation in the coherence parameters, reflecting the typical day-to-day fluctuations observed in the system. We quantify the limitation imposed by the natural cavity decoherence by tracking the state fidelity of $|\text{+}\rangle|\text{+}\rangle$ after an idle time equivalent to the duration of several CZ gates (black line). The resulting infidelity, $0.7\,\mathord{\pm}\,0.3$\% per gate, matches the expected value based on the bare lifetimes and the gate duration. 

\begin{figure}[!tbh]
    \centering
    \includegraphics{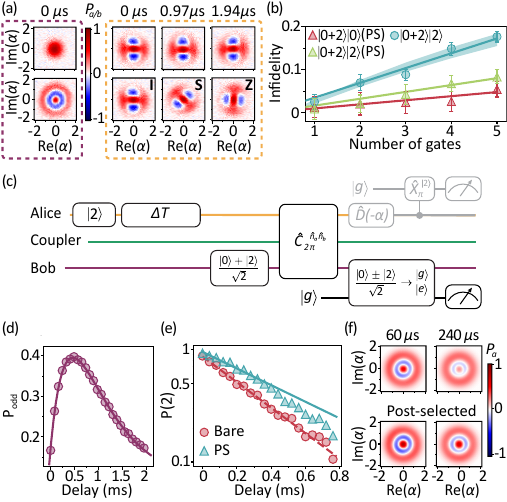}
    \caption{
    \textbf{CPHASE on a biased-erasure code.}
    (a) 
    Measured Wigner functions with Alice initialized in $(|0\rangle+|2\rangle)/\sqrt{2}$ and Bob in $|0\rangle$ or $|2\rangle$, for three different drive durations, corresponding to the controlled-I, S and Z gates. 
    (b) 
    Repeated application of $\hat C_\pi^{\hat n_a \hat n_b}$ when the control is in $|0\rangle$ (red) or $|2\rangle$ (turquoise and green). Data are offset by the initial state preparation infidelity. Dots are raw experimental data, and triangles are data post-selected on the cavities being in an even parity state after the protocol. The resulting states acquire an infidelity of 3.8$\,\mathord{\pm}\,$0.3\% (turquoise), 1.9$\,\mathord{\pm}\,$0.1\% (green), and 0.9$\,\mathord{\pm}\,$\,0.3\% (red) per gate. 
    (c) 
    Sequence to perform parity checks using the engineered cross-Kerr interaction. Alice, the storage mode, is initialized in $|2\rangle$, and allowed to decay for a variable duration up to 2\,ms. Bob, the ancillary mode, is prepared in $(|0\rangle+|2\rangle)/\sqrt{2}$. Subsequently, a $\hat C_{2\pi}^{\hat n_a \hat n_b}$ gate is applied and the state of Bob is then mapped onto the state of its auxiliary transmon with a SNAP pulse~\cite{sm}. Finally, we perform full state tomography (gray) on Alice to verify the performance of the bosonic parity check.
    (d) 
    Probability of measuring the excited state of Bob's auxiliary transmon. Open circles show the experimental data, with error bars smaller than the markers, while the solid line shows the theoretical prediction calculated using the independently measured single-photon decay time of Alice, $T_1\approx800\,\mu$s.
    (e)
    Population of Alice in Fock state $|2\rangle$ (log scale), extracted from the reconstructed density matrix, as a function of the delay time. A linear fit to the bare cavity population (red) shows a decay constant of 763$\,\mathord{\pm}\,11\,\mu$s. Post-selecting on the bosonic parity check (turquoise) and doing a linear fit to the data for delay times up to 400\,$\mu$s, where single-photon loss dominates, yields a decay constant of 1173$\,\mathord{\pm}\,41\,\mu$s.
    (f) 
    Wigner plots of the reconstructed states in Alice at two different times, with (bottom) and without (top) post-selection based on the bosonic parity check via Bob.
    }
    \label{fig:fig4_02}
\end{figure}

To demonstrate the compatibility with multi-photon encodings, we apply the engineered cross-Kerr interaction to the 0/2 subspace, which forms a biased-erasure encoding. This higher photon subspace offers two key advantages: first, the effective coupling strength increases due to the bosonic enhancement; second, standard photon-number parity checks via auxiliary transmons can be interleaved during and after the operation to detect and correct for errors, partially mitigating the drive-induced decay. 

In this subspace, we increase the detuning to $\Delta/2\pi \,\mathord{=}\,\mathord{-}10\,$MHz to account for the broader effective linewidth of the driven resonance and ensure that the residual coupler excitations are still suppressed. With this setting, we demonstrate a continuous CPHASE gate on the state $(|0\rangle+|2\rangle)/\sqrt{2}$ encoded in Alice with Bob in $|2\rangle$, indicating a CZ time of $1.94\,\mu$s, see Fig.\,\ref{fig:fig4_02}(a). Again, we extract the quality of the operation by monitoring the infidelity of the reconstructed states after repeated gate application. This yields an infidelity of $3.8\,\mathord{\pm}\,0.3\%$ per gate (turquoise line in Fig.\,\ref{fig:fig4_02}(b)). By performing a standard parity measurement using the auxiliary side transmons at the end of the protocol and post-selecting on runs in which both cavities remain in an even-parity state, the extracted infidelity is reduced to $1.9\,\mathord{\pm}\,0.1\%$ per gate (green line). For comparison, the red line corresponds to the case where the control is prepared in $|0\rangle$. In this case, we apply the same gate drive and post-select on even parity, obtaining an infidelity of $0.9\,\mathord{\pm}\,0.3\%$ per gate. The difference between the two slopes is consistent with the expected increase in decoherence when the control cavity is populated with two photons rather than zero.

Although the standard parity check using auxiliary transmons is an effective means of suppressing imperfections arising from photon loss, it necessarily entangles the microwave photons with the transmons during the process and results in excursions out of the protected bosonic code space. To fully capitalize on the advantages of bosonic QEC, it would be desirable to perform parity checks on the encoded state using purely bosonic interactions. Our engineered cross-Kerr dynamics provides the key ingredient to accomplish this goal.

In the protocol described in Fig.\,\ref{fig:fig4_02}(c), we encode a memory state $|2\rangle$ in Alice and let it relax over a variable delay time, $\Delta T$, between 0 and $2\,$ms before enacting a $\hat C_{2\pi}^{\hat n_a \hat n_b}$ gate between Alice and Bob, which is initialized in $(|0\rangle + |2\rangle)/\sqrt{2}$. This operation maps the photon number parity information of Alice onto the phase of Bob, which is then probed by an auxiliary transmon via a Selective Number-dependent Arbitrary Phase (SNAP)~\cite{heeres2015cavity} pulse that maps $(|0\rangle \pm |2\rangle)/\sqrt{2}$ to $|g\rangle/|e\rangle$. The probability of Alice being in an odd parity state after a photon loss event, $P_{\mathrm{odd}}$, which corresponds to measuring Bob's auxiliary transmon in $|e\rangle$, is shown in Fig.\,\ref{fig:fig4_02}(e). As photon loss occurs, $P_{\mathrm{odd}}$ increases and peaks around $\sim500\,\mu$s. Beyond this duration, $P_{\mathrm{odd}}$ starts to decrease as Alice decays to $|0\rangle$ after experiencing two photon loss events, recovering an even parity state. This behavior is well predicted by the analytical model (purple line), where Alice has an intrinsic single-photon lifetime of $T_1\,\mathord{\approx}\,763\,\mathord{\pm}\,11\,\mu$s. This is also consistent with our control experiment, where the decay rate of $|2\rangle$ is directly measured using Alice's auxiliary transmon (red line in Fig.\,\ref{fig:fig4_02}(e)). We consider the limit where only single-photon loss events dominate and extract an effective $T_1$ of 1173$\,\mathord{\pm}\,41\,\mu$s when we perform the bosonic parity check and post-select on Bob’s auxiliary transmon remaining in $|g\rangle$ (turquoise line in Fig.\,\ref{fig:fig4_02}(d)). We qualitatively verify this enhancement by plotting the Wigner functions of the reconstructed density matrices of Alice at two different delay times, 60$\,\mu$s and 240$\,\mu$s, in Fig.\,\ref{fig:fig4_02}(f). They indicate that our bosonic parity check effectively preserves the Wigner negativity of $|2\rangle$, which is otherwise significantly eroded by photon loss. The capability of performing such a bosonic parity check, without exposing the oscillator to any interactions outside the code space, is a crucial ingredient for effective syndrome extraction in bosonic QEC and realization of fault-tolerant algorithms.

Our implementation presents a first step towards engineering direct nonlinear couplings between oscillators and harnessing them for fault-tolerant bosonic information processing. The results reported here can be significantly improved with existing techniques to make it a valuable multi-purpose tool in bosonic cQED. In its current form, the dispersive coupling between the coupler and the oscillators leads to small variations in the effective coupling rates for different Fock states. This can be readily corrected via active cancellation techniques~\cite{rosenblum2018fault, reinhold2020error}, multi-tone drives as used in related implementations~\cite{xu2020demonstration, kim2025ultracoherent}, or by a judicious choice of operating point that minimizes the resulting inhomogeneity across Fock states~\cite{susan}. In addition, a promising alternative is to replace the transmon coupler with one that does not impart any dispersive shifts, such as the Superconducting Nonlinear Asymmetric Inductive eLement (SNAIL)~\cite{frattini20173, sm}. 

The primary limitation of the current gate fidelity is the drive-induced photon loss of the oscillators and the dressed dephasing arising from the interplay between the enhanced photon loss and the engineered cross-Kerr~\cite{sm}. This is a known challenge in four-wave mixing architectures operating under strong drives~\cite{carde2025flux, zhang2019engineering} and an active area of ongoing research~\cite{dai2026characterization}. One potential mitigation strategy is to use the SNAIL or other more tailored nonlinear couplers~\cite{maiti2025linear}, which reduce the four-wave-mixing terms and suppress the decoherence effects stemming from these spurious dynamics. Alternatively, we can also leverage broadband flux-tunability to dynamically tune the coupler to the operating point where the Raman condition is satisfied, eliminating drive-induced decoherence entirely~\cite{valadares2026flux}. 

In conclusion, our work demonstrates a dynamically activated cross-Kerr coupling between microwave photons hosted in two superconducting cavities without exciting the nonlinear coupler, with an interaction strength two orders of magnitude stronger than the residual always-on coupling. We leverage this direct nonlinear coupling between the oscillators to implement a continuous CPHASE gate in the 0/1 and 0/2 photon subspaces, with an average gate infidelity below 5\%. Furthermore, we use the cross-Kerr dynamics to perform a direct bosonic parity check, where photon loss of a storage cavity is detected via another ancillary oscillator mode. This suppresses any ancilla-induced decoherence mechanisms and ensures that the storage oscillator remains entirely within the protected bosonic code space during the syndrome measurement. These results present the first direct cross-Kerr interaction between two harmonic oscillators realized without explicitly populating nonlinear coupling elements. Importantly, the resulting CPHASE gate is natively compatible with the full family of rotationally-symmetric codes and provides a critical ingredient for fault-tolerant QEC protocols~\cite{grimsmo2020quantum}. Thus, our technique marks a promising starting point for exploring the rich physics of nonlinear quantum optics and implementing robust quantum information processing using bosonic elements. \\

\noindent\textbf{Acknowledgments} We acknowledge the funding support from the Singapore Ministry of Education (MOE-T2EP50222-0017) and The University of Sydney - National University of Singapore 2026 Ignition Grants. A.C., A.K., S.Q., acknowledge the support of the Singapore National Quantum Scholarship Scheme (NQSS). We thank Mr.\,Nixon Yang, Mr.\,Juncheng Man, and Dr.\,Yao Lu for technical inputs during the project development.

\clearpage
\onecolumngrid

\begin{center}
{\Large\bfseries Supplementary information: A direct controlled-phase gate between microwave photons}

\end{center}

\vspace{0.5cm}

\setcounter{section}{0}
\setcounter{equation}{0}
\setcounter{figure}{0}
\setcounter{table}{0}
\renewcommand{\thesection}{S\arabic{section}}
\renewcommand{\theequation}{S\arabic{equation}}
\renewcommand{\thefigure}{S\arabic{figure}}
\renewcommand{\thetable}{S\arabic{table}}

\twocolumngrid

\section{Experimental device and system parameters}

The experimental device used in this work consists of two oscillators in the form of the standard three-dimensional stub cavities, Alice and Bob, machined out of high-purity (4N6) aluminum. They are coaxial $\lambda$/4-resonators with cut-off frequencies around $f_\text{cut}\,\mathord{\sim}\,600$\,MHz and stub lengths of 16.5\,mm and 15.3\,mm, which correspond to frequencies of $4.12\,$GHz and $4.41\,$GHz, respectively. The external surface layer ($\mathord{\sim}\,$0.15\,mm) of the device has been chemically removed with aluminium etchant type A to reduce fabrication imperfections, achieving Q factors of $\mathord{\sim}$\,2$\cdot10^7$ and $\mathord{\sim}$\,3 $\cdot10^7$, respectively. Each oscillator is capacitively coupled to a standard auxiliary transmon that is used for state preparation and readout. The electromagnetic field distributions were simulated using Ansys HFSS, and the Hamiltonian parameters were determined using the energy participation ratio (EPR) approach provided by the python package pyEPR~\cite{minev2021energy}. The Hamiltonian of both cavities and their side chips has the form
 \begin{equation}
\begin{split}
H_i/\hbar &=
\omega_i i^\dagger i
+ \omega_{q_i} \hat q_i^\dagger \hat q_i
+ \omega_{rr_i} \hat r_i^\dagger\hat  r_i
\\
& -\frac{\alpha_{q_i}}{2} \hat q_i^\dagger \hat q_i^\dagger \hat q_i \hat q_i
- \frac{K_i}{2} \hat i^\dagger \hat i^\dagger \hat i \hat i
\\
& - \chi_{i\,q_i}\, \hat i^\dagger \hat i\,\hat q_i^\dagger \hat q_i 
- \chi_{q_i\,rr_i}\, \hat q_i^\dagger \hat q_i\, \hat r_i^\dagger \hat r_i,
\end{split}
\label{Eq: side Hams}
\end{equation}
where $\hat i=\hat a, \hat b$, for both oscillators. All the parameters of the Hamiltonian and their physical meaning are summarized in table~\ref{tab: Subsystems}.

\begin{table}[htbp]
  \centering
  \begin{tabular}{ccc}
    \hline
    Parameter & Description & Value \\
    \hline
    $\omega_a/2\pi$ & Alice frequency & 4.12\,GHz \\
    $\omega_{q_a}/2\pi$ & Transmon A frequency & 5.65\,GHz \\
    $\omega_{rr_a}/2\pi$ & Resonator A frequency & 7.20\,GHz \\
    $\alpha_{q_a}/2\pi$ & Transmon A anharmonicity & 118.3\,MHz \\
    $K_a/2\pi$ & Alice self-Kerr & 1.3\,kHz \\
    $\chi_{a\,q_a}/2\pi$ & Transmon A - Alice disp. shift & 0.77\,MHz \\
    $\chi_{q_a\,rr_a}/2\pi$ & Transmon A - Resonator A disp. shift & 0.6\,MHz \\
    \hline

    $\omega_b/2\pi$ & Bob frequency & 4.41\,GHz \\
    $\omega_{q_b}/2\pi$ & Transmon B frequency & 5.18\,GHz \\
    $\omega_{rr_b}/2\pi$ & Resonator B frequency & 7.32\,GHz \\
    $\alpha_{q_b}/2\pi$ & Transmon B anharmonicity & 137.9\,MHz \\
    $K_b/2\pi$ & Bob self-Kerr & 2.4\,kHz \\
    $\chi_{b\,q_b}/2\pi$ & Transmon B - Bob disp. shift & 2.16\,MHz \\
    $\chi_{q_b\,rr_b}/2\pi$ & Transmon B - Resonator B disp. shift & 0.45\,MHz \\
    \hline

    $\omega_c/2\pi$ & Coupler frequency & 5.22\,GHz \\
    $\alpha_{c}/2\pi$ & Coupler anharmonicity & 196\,MHz \\
    $\chi_{a}/2\pi$ & Coupler - Alice disp. shift & 0.8\,MHz \\
    $\chi_{b}/2\pi$ & Coupler - Bob disp. shift & 0.5\,MHz \\
    $\omega_{rr_c}/2\pi$ & Coupler readout frequency & 7.49\,GHz \\
    $\chi_{c\,rr_c}/2\pi$ & Coupler - readout disp. shift & 0.8\,MHz \\
    \hline
  \end{tabular}
  \caption{\textbf{Hamiltonian parameters of the device.} The parameters of the coupler and its readout resonator correspond to the operating flux-bias point used in Figures~3 and 4 of the main text.}
  \label{tab: Subsystems}
\end{table}

To allow for fast initial state preparation and fast tomography, both oscillators couple to the respective auxiliary transmons in the strong dispersive coupling regime. The strength of this dispersive coupling is chosen such that the inherited self-Kerr non-linearity of the cavity modes, $K_{a/b}$, is in an acceptable range of 1-3\,kHz. All the states used in the main text are prepared using numerically optimized Gradient Ascent Pulse Engineering (GRAPE) pulses~\cite{heeres2017implementing} of $2\,\mu$s length. Tomography is performed either by sampling the Wigner function and measuring the parity of the cavity state at different points in phase space using a standard Ramsey technique, or by using the Optimized Reconstruction via Excitation Number Sampling (ORENS) technique~\cite{krisnanda2025demonstrating}. See Section~\ref{Section: Quantum state reconstruction} for further details.

\subsection{Coupler frequency arrangement}

Both oscillators are capacitively coupled to a flux-tunable asymmetric Superconducting Quantum Interference Device (SQUID)~\cite{hutchings2017tunable} that acts as the coupler in these experiments. While flux-tunability is not strictly necessary for our protocol, it provides valuable flexibility by allowing us to avoid frequency configurations that may introduce unintended higher-order transitions and to explore different coupler frequencies within a single cooldown.

As the auxiliary transmons do not participate in the main dynamics of the engineered cross-Kerr coupling, we now focus only on the Hamiltonian of the coupler and both oscillators, given by: 
\begin{equation}
\begin{split}
H_c/\hbar &= \omega_a \hat a^\dagger \hat a - \frac{K_a}{2} \hat a^\dagger \hat a^\dagger \hat a \hat a \\
&+ \omega_b \hat b^\dagger \hat b - \frac{K_b}{2} \hat b^\dagger \hat b^\dagger \hat b \hat b \\
&+ \omega_{c} \hat c^\dagger \hat c - \frac{\alpha_{c}}{2} \hat c^\dagger \hat c^\dagger \hat c \hat c \\
&- \chi_{a} \hat a^\dagger \hat a \hat c^\dagger \hat c - \chi_{b} \hat b^\dagger \hat b \hat c^\dagger \hat c - \frac{K_{ab}}{2} \hat a^\dagger \hat a \hat b^\dagger \hat b\\
&+ \omega_{rr_C} \hat r^\dagger \hat r - \chi_{c\,rr_c} \hat c^\dagger \hat c \hat r^\dagger \hat r,
\end{split}
\end{equation}
where the experimental parameters associated with each mode used in Figures 3 and 4 in the main text are summarized in table~\ref{tab: Subsystems}.

To engineer the Raman-assisted interaction described in the main text, we parametrically drive the coupler at a frequency $\omega_d \,\mathord{\approx}\,\omega_c \,\mathord{\pm}\, |\omega_b - \omega_a|$. Drive frequencies closer to the coupler frequency translate into a higher effective drive strength $\xi$. Thus, we design the oscillator frequencies $\omega_a$ and $\omega_b$ at close spectral proximity $\left(\omega_b - \omega_a\right)/2\pi\, \mathord{\approx} \,284\,$MHz. Since this detuning is comparable to the coupler anharmonicity, we chose to drive the $a^\dagger b |g\rangle\langle e| +h.c.$ rather than the $a b^\dagger |g\rangle\langle e| +h.c.$ transition, in order to avoid driving the coupler in its straddling regime. Both processes are suitable to generate the desired cross-Kerr dynamics.

Placing the coupler frequency above the cavities allows us to tune the coupling regime of the system from essentially decoupled to almost resonant, as we increase the biasing flux. Coupler frequencies close to the cavities result in an increased always-on natural cross-Kerr, decreasing the on/off ratio of the gate. On the other hand, when the coupler is close to its upper sweet spot, we observe a drop in both oscillators dephasing times. We attribute this to photon shot noise caused by thermal excitations in the coupler and readout mode. Since the strength of the coherent exchange interaction $g_1$ is proportional to the dispersive shift of both cavities, we must consider the tradeoff between gate speed and on-off ratio. Balancing these considerations, we typically operate at the coupler frequency range of $5.1\pm0.2\,$GHz, where the natural cross-Kerr $K_{ab}$ is below the kHz level.

\subsection{Coupler flux delivery}

Introducing magnetic flux while preserving the coherence of the cavities in a superconducting package is a challenging task~\cite{valadares2024demand}. Here, we follow the approach in~\cite{chapman2023high}, where an on-chip superconducting pick-up loop~\cite{zimmerman1971sensitivity} transfers flux from a nearby magnetic coil to the SQUID. The loop works as a transformer, where the magnetic flux originating from a small superconducting coil threads the pick-up loop, inducing a screening current which then biases the SQUID loop. 

The magnetic coil generating the initial magnetic field consists of $\mathord{\sim}$\,500 turns of NbTi superconducting wire (SC-54S43-0.178\,mm wire from Supercon) coiled around an OFHC copper spool of 1\,cm length and 4.5\,mm diameter. The terminals of the coil are soldered to the DC lines at the mixing chamber (MXC) stage of the dilution refrigerator. The spool is thermally anchored to the lid of the cavities, which is also made of OFHC copper, and is thermally anchored directly to the MXC stage using a copper braid. We observed a maximum MXC stage temperature of 26\,mK when passing a current of 60\,mA.

\begin{figure}[t!]
    \centering
    \includegraphics{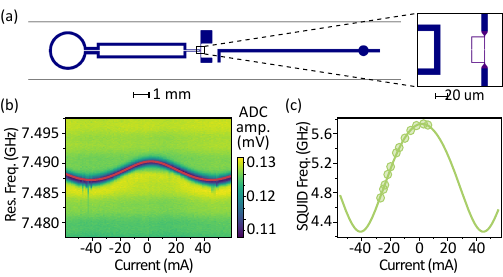}
    \caption{\textbf{SQUID coupler tunability.} 
    (a) Layout of the SQUID coupler chip, with the pick-up loop transformer (left), the SQUID (middle), and its readout resonator (right).
    (b) Readout frequency as a function of the current applied to bias the SQUID. Solid line shows the fit used to extract the current-flux relation. 
    (c) SQUID frequency as a function of current. Solid line shows the fitted frequency.}
    \label{fig:flux}
\end{figure}

The design of the SQUID chip is shown in Fig.\,\ref{fig:flux}(a). The geometry of the pick-up loop was optimized using Ansys Maxwell to maximize the flux delivered to the device while minimizing the current required in the coil. Further, we also verify that the design preserves the Q factors of both cavities and does not introduce undesired modes in the system. The final configuration consists of 4 distinct sections: (i) a circular region optimized to only collect the magnetic flux piercing the chip; (ii) a constriction to avoid collecting the returning field; (iii) the long sections of the transformer, spaced out to reduce the loop self-inductance; and (iv) a narrowed section near the SQUID loop to avoid interfering with the junction fabrication during the lithography step. 

The length of the pick-up loop is chosen to balance two effects. First, it needs to be long enough such that the copper coil and spool are spatially separated from the cavities, both for ease of assembly and to prevent the electric field of the cavity TEM modes from leaking out through the aperture of the coil, which would degrade the cavity lifetimes. Second, the parasitic resonant modes of the transformer, currently at 6.2\,GHz and 8.1\,GHz, decrease in frequency as its length increases, potentially interfering with other modes in the system. Finally, the loop formed by both junctions of the SQUID, which has an area of 800\,$\mu\text{m}^2$, is placed at a distance of $\mathord{\sim}$\,50\,$\mu$m from the flux transformer, see inset in Fig.\,\ref{fig:flux}(a). The transformer has a width of 10\,$\mu$m at its narrowest point, ensuring that it remains superconducting even with large induced currents. 

To mitigate the dephasing of the flux-tunable coupler, we compromise on the frequency tunability range by making both junctions asymmetric. Given the junctions Josephson energies $E_{J_1}$ and $E_{J_2}$, the effective Josephson energy of the SQUID is
\begin{equation}
    E_J \left(\Phi_\text{ext}\right)=E_{J,\Sigma} \cos\left(\pi\frac{ \Phi_\text{ext}}{\Phi_0}\right)\sqrt{1+\frac{ E_{J,\Delta}}{ E_{J,\Sigma}}\tan^2
    \left(\pi\frac{\Phi_\text{ext}}{\Phi_0}\right)},
\end{equation}
where $E_{J,\Sigma}\,\mathord{=}\, E_{J_1}+E_{J_2}$ and $E_{J,\Delta}\,\mathord{=}\, E_{J_1}-E_{J_2}$. The measured frequency of the asymmetric SQUID as a function of the applied current on the coil is shown in Fig.\,\ref{fig:flux}(c). Due to the increased attenuation of the drive line at lower frequencies and the reduced dispersive coupling to its readout resonator, we cannot measure the coupler frequency for values below 4.7\,GHz. However, by fitting the SQUID readout resonator frequency as a function of the applied current through the coil, we extract the relation between the current and the flux in the SQUID loop, with 42.89\,mA corresponding to half a flux quantum, see Fig.\,\ref{fig:flux}(b). Finally, to extract the whole tunability range of the SQUID and the individual Josephson energies of both junctions, we consider the coupler-resonator Hamiltonian, 
\begin{equation}
\begin{split}
    \hat H_c/\hbar\,&=\,4E_C \hat n^2 - E_J \left(\Phi_\text{ext}\right) \cos \hat\phi + \omega_{rr_c} \hat r^\dagger \hat r \\
    &+ g (\hat c-\hat c^\dagger)(\hat r-\hat r^\dagger)
\end{split}
\end{equation}
where $E_C$ is the charging energy of the SQUID, $\hat n $ and $\hat \phi$ are the charge and phase operators, respectively, $\omega_{rr_c}$ is the readout resonator frequency and $g$ is the SQUID-resonator coupling strength. Fitting the frequencies of these two modes as a function of flux, constrained by the measured anharmonicity and the current-flux relation, we obtain the parameters shown in Table~\ref{tab: SQUID params}. This allows us to determine the full SQUID tunability, spanning 4.3\,GHz to 5.7\,GHz, see solid line in Fig.\,\ref{fig:flux}(c), as well as the SQUID asymmetry, $E_{J_1}/E_{J_2}\, \mathord{\approx} \,3.7$, which closely matches the designed asymmetry during fabrication. These are the parameters used in Floquet simulation in Section~\ref{Section: Floquet simulations}.

\begin{table}[h]
\centering
\begin{tabular}{ccc}
\hline
Parameter & Value & Description \\
\hline
$E_{J_1}$ & $19.0 \pm 0.1$\,GHz & Josephson energy \\
$E_{J_2}$ & $5.2 \pm 0.1$\,GHz & Josephson energy \\
$E_{C}$ & $181.4 \pm 0.1$\,MHz& Charging energy \\
$g$ & $0.10\pm 0.02$\,GHz & Coupling constant \\
$\omega_{r_0}/2\pi$ & $7.484\pm 0.002$\,GHz & Resonator frequency \\
\hline
\end{tabular}
\caption{\textbf{Hamiltonian parameters of the SQUID coupler}. Parameters were fitted to reproduce the resonator and coupler flux tunability as a function of the applied current, with the additional constraint of reproducing the experimental anharmonicity of the coupler.}
\label{tab: SQUID params}
\end{table}

\subsection{System coherences}
\label{subsec:System Coherences}
Each transmon and its readout resonator are addressed through a shared coupling port, whose position is optimized in Ansys HFSS to suppress radiative decay of the transmons into the measurement lines, thereby achieving intrinsic Purcell filtering~\cite{sunada2022fast}. This approach preserves transmon coherence without the need for on-chip Purcell filters, reducing chip size, mitigating susceptibility to vibrations, and enabling a more compact device. The coherence times of both side transmons and the SQUID coupler are shown in Table~\ref{tab:System Coherences A and B}. The coupler coherence times depend strongly on the it frequency configuration, with $T_1$ times ranging from $50\,\mu$s at the lowest-measured frequency to $10\,\mu$s at its upper sweet spot, where the coupler hybridizes more strongly with its readout resonator. On the other hand, the dephasing times range from up to $20\,\mu$s at the flux sweet spot to $1\,\mu$s at the most sensitive location. The coupler coherence times in the configuration used to obtain the final data for Figures 3 and 4 in the main text are shown in Table~\ref{tab:System Coherences A and B}. Importantly, although the coupler is operated near its most flux-sensitive point, its intrinsic decoherence does not significantly affect the quality of the gate, since the Raman-assisted interaction is mediated through virtual processes that avoid populating its excited states. 

\begin{table}[htbp]
  \centering
  \begin{tabular}{cc|cc|cc}
    \hline
    Param. & Value ($\mu$s) & Param. & Value ($\mu$s) & Param. & Value ($\mu$s) \\
    \hline
    $T_{1q_a}$ & 25 & $T_{1q_b}$ & ~50-60 & $T_{1c}$ & 50\\
    $T_{2q_a}^*$ & 18 & $T_{2q_b}^*$ & ~40-50 & $T_{2c}^*$ & 2.5 \\
    $T_{2q_a}^E$ & 22& $T_{2q_b}^E$ & 50 & $T_{2c}^E$ & 10 \\
    $T_{1a}$& 700-900 & $T_{1b}$ & 800-1000 \\
    $T_{2a}$  & 500-600 & $T_{2b}$ & 700-900  \\
    \hline
  \end{tabular}
  \caption{\textbf{Device coherence times.} Coherence times for the coupler are quoted at the flux point used in the main text. Transmon $T_2^*$ were measured with a standard Ramsey technique, while $T_2^E$ times were obtained via a Hahn-echo experiment. Oscillator $T_1$ times were measured by fitting the decay of both coherent states and single photons. Oscillator $T_2$ times were measured through a Ramsey experiment using SNAP pulses.}
  \label{tab:System Coherences A and B}
\end{table}

In contrast, cavity decoherence directly limits the achievable quality of the engineered cross-Kerr interaction. The cavities are designed to be internal Q limited, with coupling ports designed to have coupling quality factors on the order of $\mathord{\sim}\,10^9$. Good care was also taken to minimize the cavity photon field from leaking into the coil used to bias the SQUID. We measured cavity $T_1$ times both by initializing a coherent state and measuring its characteristic decay, and by loading a photon in the cavity using SNAP pulses and fitting an exponential decay. Both methods give consistent results with average single-photon lifetimes of $800\,\mu$s and $900\,\mu$s for Alice and Bob, respectively. The cavity $T_2$ times were measured using a standard Ramsey technique, where they are initialized in $(|0\rangle + |1\rangle)/\sqrt{2}$ using SNAP pulses and, after a variable wait time, they are displaced by $D(\alpha\,\mathord{=}\,-0.8e^{i\omega t})$. This displacement maximizes the overlap with vacuum for $|0\rangle + |1\rangle)$ and minimizes it for $|0\rangle - |1\rangle)$, and $\omega$ is a fixed virtual detuning that facilitates extracting the $T_2$ dephasing time by fitting the data to an exponentially decaying sinusoid.

\subsection{Auxiliary transmon readout}
\label{Section: readout}
Both the auxiliary transmons and the SQUID coupler are measured using Cavity Level Excitation and Reset (CLEAR) readout pulses~\cite{mcclure2016rapid} applied to their respective readout resonators. In contrast to a conventional constant-amplitude readout pulse, the CLEAR protocol employs a sequence of five piecewise-constant segments: two ring-up segments, one steady-state segment, and two ring-down segments. This pulse shaping offers high-fidelity state discrimination while significantly reducing the residual photon population after the measurement pulse. Compared to standard square pulses, CLEAR pulses achieve faster resonator reset and therefore support reduced measurement-induced backaction and shorter experimental cycle times. The amplitudes and duration of the individual CLEAR drive segments are optimized running an algorithm that minimizes the population of the resonator after the CLEAR pulse given the decay time of the resonator $1/\kappa$ and the dispersive shift to the transmon $\chi$. 

In addition to CLEAR pulse shaping, our readout is typically optimized to differentiate between states $|g\rangle$ and $|f\rangle$, which improves the separation of the histogram counts for the same integration time compared to discriminating between $|g\rangle$ and $|e\rangle$, as $\chi_{gf} \,\mathord{\approx}\, 2\,\chi_{ge}$. After each experimental sequence we map the $|e\rangle$ population to the $|f\rangle$ state using a $\sigma=24\,$ns gaussian pulse. The resulting confusion matrices for both auxiliary transmons used in the tomography and state preparation steps are:

\[
M_{\mathrm{RO},A} =
\begin{pmatrix}
P(g|g)=97.5\%& P(e|g)=2.5\% \\
P(g|e)=2.8\%& P(e|e)=97.2\%
\end{pmatrix}
\]

\[
M_{\mathrm{RO}, B} =
\begin{pmatrix}
P(g|g)=95.9\%& P(e|g)=4.1\%\\
P(g|e)=6.4 \%& P(e|e)=93.6\%
\end{pmatrix}
\]

\subsection{Chip fabrication}
The ancillary transmon chips and the SQUID coupler chip are fabricated in aluminum on a sapphire substrate. A HEMEX sapphire wafer is cleaned in a 2:1 piranha solution for 20 minutes and rinsed in de-ionized (DI) water for another 20 minutes. It is then quickly rinsed in methanol and blown-dry with nitrogen. The wafer is coated with 700\,nm of MMA and 200\,nm of PMMA resist by spinning it for 100 seconds at 2000\,rpm and baking it for 5 minutes at 200$^{\circ}$C. A $\sim\,$10-nm discharge gold layer is sputtered in a Cressington sputterer at 30\,mA for 30 seconds. The design is patterned using a Raith electron-beam lithography machine. The gold layer is removed in a KI solution and rinsed in DI water, prior to developing the resist in a 3:1 mixture of DI water and isopropanol at 6$^{\circ}$C for 2 minutes. The wafer is then loaded into an Angstrom Engineering double-angle evaporator and pumped to $10^{-8}\,$mbar, after which the resist is ion milled in a mixture of 85\% O2 and 15\% Argon, at 400\,V for 15\,s at -20$^{\circ}$ and 15\,s at 20$^{\circ}$ to clean any resist residues. We deposit two aluminum layers of 20\,nm and 30\,nm thickness at -20$^{\circ}$ and +20$^{\circ}$, respectively, separated by an oxidation step with a mixture of 85\% O2 and 15\% Argon at 20\,mbar for 20 minutes. Before unloading the wafer, a final capping layer with the same oxidation conditions is performed to grow a more controlled surface oxide. The remaining resist is then lifted off in NMP at 90$^{\circ}$C for 3 hours, then rinsed in acetone and methanol. Finally, a protective layer of AZ1512 photoresist is spun at 2000\,rpm and baked at 80$^{\circ}$C for 1 minute, and the individual chips are diced on an Accretech machine with a resin blade at 15000\,rpm. The chips are finally cleaned in NMP, acetone and methanol, blown-dry and inserted into the waveguides using an aluminum clamps with indium wire to improve thermalization.

\section{SNAP pulses}
\label{Section: SNAP pulses}

We use Selective Number-dependent Arbitrary Phase (SNAP) pulses~\cite{heeres2015cavity} to prepare the oscillators in Fock state $|1\rangle$
and the superposition state $(|0\rangle + |1\rangle)/\sqrt{2}$, to measure the single-photon decay rate and dephasing rate of the oscillators, respectively. The SNAP pulses consist of an initial displacement of the oscillators by $\alpha_1$, a phase, $\theta_n$, imparted on the transmon state conditioned on the oscillator having $n$ photons, and a final displacement of the oscillators by $\alpha_2$. All parameters used can be found in Table~\ref{tab: SNAP}. Oscillator displacements were performed using 100\,ns constant pulses with 32\,ns cosine ramps at the resonant frequency of the oscillators. The conditional $\theta_n\,\mathord{=}\,\pi$ phases were realized by driving the transmons through a full $2\pi$-rotation around the Bloch sphere, using gaussian pulses of $\sigma_a\,\mathord{=}\,960\,$ns and $\sigma_b\,\mathord{=}\,400\,$ns, applied at the resonance frequency of each transmon conditioned on the oscillator containing $n$ photons.

\begin{table}[h]
\centering
\renewcommand{\arraystretch}{1.4}
\begin{tabular}{l@{\hspace{1em}}@{\hspace{0.5em}}c@{\hspace{1em}}c@{\hspace{1em}}c@{\hspace{1em}}c}
\toprule
\textbf{Transition} & $\alpha_1$ & $\theta_0$ & $\theta_1$ & $\alpha_2$ \\
\midrule
$|0\rangle \rightarrow |1\rangle$                       & 1.14 & $\pi$ & $0$   & $-0.58$ \\
$|0\rangle \rightarrow \frac{|0\rangle+|1\rangle}{\sqrt{2}}$          & 0.56 & $\pi$ & $0$   & $-0.24$ \\
$\frac{|0\rangle-|2\rangle}{\sqrt{2}} \rightarrow |0\rangle$          & -0.35 & $\pi$ & $\pi$ & 1.04 \\
\bottomrule
\end{tabular}
\caption{\textbf{Different SNAP parameters used in the main text}. First row corresponds to the creation of a single photon in the oscillators, used to measure their decay time. Second row corresponds to the creation of a superposition state in the oscillators, used to measure their dephasing time in a Ramsey experiment. Last row corresponds to the pulse used in the direct cavity-cavity parity mapping, shown in Fig.\,4(c) of the main text.}
\label{tab: SNAP}
\end{table}

In the direct parity mapping protocol described in Fig.\,4(c) in the main text, we use a SNAP pulse to map the even and odd superposition states of Bob to
\begin{equation*}
\begin{split}
     \frac{|0\rangle - |2\rangle}{\sqrt{2}} & \rightarrow |0\rangle\\
     \frac{|0\rangle + |2\rangle}{\sqrt{2}} & \rightarrow  0.9\,|1\rangle + 0.4\,|2\rangle + \dots
\end{split}
\end{equation*}
A selective $\pi$-pulse to Bob's auxiliary transmon conditioned on Bob being in vacuum hence maps $(|0\rangle - |2\rangle)/\sqrt{2}$ and $(|0\rangle + |2\rangle)/\sqrt{2}$ to $|e\rangle$ and $|g\rangle$, respectively. The displacement amplitudes used to implement this transformation were obtained via a Nelder–Mead optimization that maximizes the probability of the final state being either in vacuum or not vacuum. The parameters of the SNAP sequence are listed in the last row of Table~\ref{tab: SNAP}.

\section{Wiring diagram}
The microwave signals to drive the oscillators, auxiliary transmons, coupler, and readout resonators of our device are directly generated from digital waveforms via Direct-Digital-Synthesis (DDS) by a Digital-to-Analog Converter (DAC) of a Quantum Machines OPX1000 Field-Programmable Gate Array (FPGA). We also place band-pass filters at the output line of the OPX1000, which increased the transmon coherence times significantly. Additionally, a RF switch (HMC-C058) placed on the coupler line provides an additional layer of noise protection. Importantly, most of the coupler drive line attenuation at the MXC stage is replaced by a Mini-circuits 5500+ low-pass filter and Mini-circuits 5500+ high-pass filter, which provide -30\,dB attenuation at 5.0\,GHz but only -10\,dB at 5.5\,GHz to be able to drive the coupler strongly, while still providing at least 60\,dB attenuation at the oscillator frequencies and 35\,dB at the resonator frequency. 

Readout is performed using the standard reflection technique. The reflected signal, amplified by a High-Electron mobility transistor (HEMT - LNF-LNC48C) as well as a ZVA-183-S+ room temperature amplifier, is directly sampled by a high-speed Analog-to-Digital Converter (ADC) of the OPX1000 FPGA. The digital signal is then downconverted and separated into I and Q components. The DC current sent to the coil is generated in a YOKOGAWA GS200 DC voltage/current source. To shield our device from external electromagnetic noise it is placed in a standard Cryoperm shield and anchored directly at the MXC plate. The schematic of the wiring can be seen below in Fig.~\ref{fig:wiring}

\begin{figure*}[!tbh]
    \centering
    \includegraphics{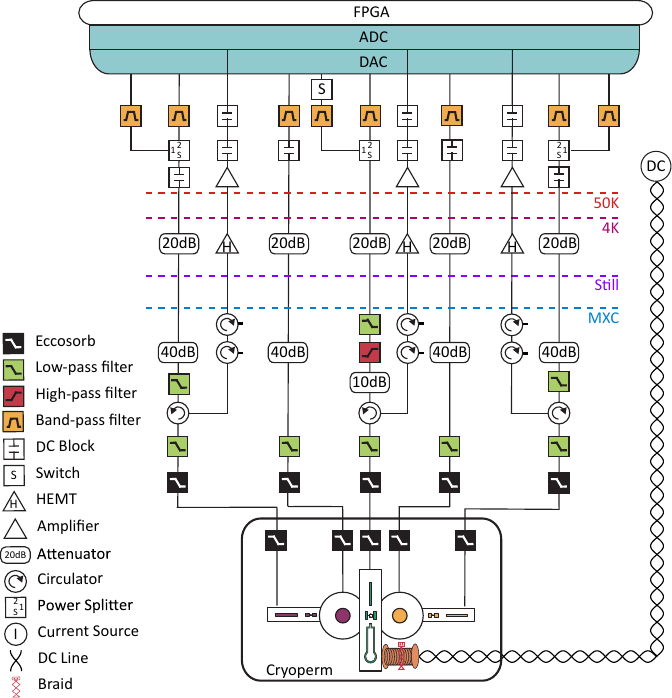}
    \caption{\textbf{Experimental wiring setup}. Schematic of the RF components and connections at room temperature and inside the Bluefors dilution refrigerator. Most of the attenuation for the coupler drive line is replaced by a low-pass filter in series with a high-pass filter that act as a 5-7\,GHz band-pass filter.}
    \label{fig:wiring}
\end{figure*}

\section{Analytical derivation of the engineered dynamics}~\label{Section: SWT}

The Hamiltonian of the two oscillators and the SQUID coupler is
\begin{equation}
\begin{split}
    H(t)/\hbar &= \tilde\omega_a~\hat{a}^\dagger \hat{a} + \tilde\omega_b~\hat{b}^\dagger \hat{b} \\ &+\tilde\omega_c~\hat{c}^\dagger \hat{c}-E_J(\Phi_\text{ext}) \left( \cos \hat\phi + \frac{1}{2!} \hat\phi^2 \right) \\
    &+\epsilon(t) (\hat{c}+\hat{c}^\dagger),
\end{split}
\label{Eq: H(t)}
\end{equation}
where $\tilde\omega_a$ and $\tilde\omega_b$  are the bare (uncoupled) oscillator frequencies, $\tilde\omega_c\,\mathord{=}\,\sqrt{8 E_J(\Phi_\text{ext}) E_C}$ is the Josephson plasma frequency of the coupler, $\epsilon(t)$ is the time-dependent drive on the coupler and 
$$\hat \phi=\phi_c ~(\hat{c}+\hat{c}^\dagger) + \phi_a ~(\hat a+\hat a^\dagger) + \phi_b ~(\hat b+\hat b^\dagger)$$ is the quantized phase of the coupler, that considers the participation of both oscillator fields in the junction.

The zero-point fluctuations of the coupler in the junction is defined as $\phi_c\,\mathord{=}\,\left( 2 E_C/E_J(\Phi_\text{ext})\right)^{1/4}$, while the zero-point fluctuations of the oscillators in the junction are approximated by $\phi_a \,\mathord{\approx}\, \frac{g_a}{\Delta_a} \phi_c$ and $\phi_b\, \mathord{\approx}\, \frac{g_b}{\Delta_b} \phi_c$. Considering a drive of the form $\epsilon(t)\,\mathord{=}\,2\varepsilon \cos \left(\omega_d ~t+\phi_d \right)$, we can go to the displaced frame $\hat{c} \rightarrow \hat{c} + \xi e^{-i\omega_d t}$ with $\xi\,\mathord{=}\,\varepsilon/(\omega_d -\tilde\omega_c)$, where the drive term is absorbed into the phase variable through $$\hat \phi=\phi_c ~(\hat{c}+\hat{c}^\dagger + \xi e^{-i\omega_d t} + \xi^* e^{i\omega_d t}) + \phi_a ~(\hat a+\hat a^\dagger) + \phi_b ~(\hat b+\hat b^\dagger).$$ By expanding the cosine in Eq.\,\ref{Eq: H(t)} up to fourth order in $\hat \phi$, and keeping all the terms that are naturally on-resonance, we obtain the full Hamiltonian of the system
\begin{equation}
\begin{split}
\hat H_0 &=  \omega_a~\hat{a}^\dagger \hat{a} -\frac{K_a}{2}\hat{a}^\dagger\hat{a}^\dagger\hat{a}\hat{a} \\ 
        & + \omega_b~\hat{b}^\dagger \hat{b} -\frac{K_b} {2}\hat{b}^\dagger\hat{b}^\dagger\hat{b}\hat{b} \\ 
        & + \omega_c~\hat{c}^\dagger \hat{c} -\frac{\alpha}{2}\hat{c}^\dagger\hat{c}^\dagger\hat{c}\hat{c} \\
        & - \chi_{a}~\hat{a}^\dagger\hat{a} ~\hat{c}^\dagger\hat{c} -\chi_{b}~\hat{b}^\dagger\hat{b} ~\hat{c}^\dagger\hat{c} -K_{ab}~\hat{a}^\dagger\hat{a} ~\hat{b}^\dagger\hat{b},
\end{split}
\end{equation}
where $K_a$ and $K_b$ are the cavity self-Kerrs inherited from the coupler non-linearity, 
$\alpha$ is the coupler non-linearity (anharmonicity), $\chi_{a}$ and  $\chi_{b}$ are the coupler-cavity cross-Kerrs (dispersive shifts), and $K_{ab}$ is the residual always-on cavity-cavity cross-Kerr. All these terms produce Lamb and Stark shifts on the bare mode frequencies, which get renormalized as
\begin{align*}
    \omega_a &= \tilde\omega_a - \frac{\chi_{a}}{2} - K_{a} - \frac{K_{ab}}{2} - \chi_{a}|\xi|^2\\
    \omega_b &= \tilde\omega_b - \frac{\chi_{b}}{2} - K_{b} - \frac{K_{ab}}{2} - \chi_{b}|\xi|^2\\
    \omega_c &= \tilde\omega_c - \alpha - \frac{\chi_{a}}{2} -  \frac{\chi_{b}}{2} -2\alpha|\xi|^2,
\end{align*}
where the last term of each equation corresponds to the drive-induced AC Stark shifts.

\begin{figure}[tbh!]
\centering
\includegraphics {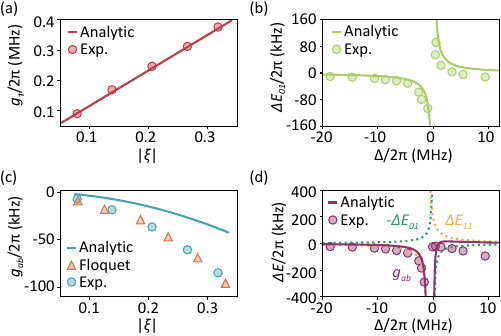}
\caption{\textbf{Analytical predictions based on the SWT.} 
(a) 
Coherent exchange rate, $g_1$, as a function of the effective drive strength, $|\xi|$ (red circles). The analytical prediction of Eq.\,\ref{Eq: g1 derivation} is in close agreement (solid line).
(b) 
Phase rate acquired by the state $|\mathord{+}\rangle|0\rangle|g\rangle$ (green circles) as a function of the detuning $\Delta$ to the coherent exchange condition. Analytics correspond to the prediction from Eq.\,\ref{Eq: 01 rate} (solid line).
(c) 
Cross-Kerr strength (blue circles) measured at a fixed $\Delta\,\mathord{=}\,\mathord{-}5$\,MHz, as a function of drive power. Analytical prediction (solid line) based on Eq.\,\ref{Eq: cross_kerr_after_SWT} under-predicts the strength by a factor of 2. However, a full Floquet simulation (orange triangles) is in very good agreement with the data.
(d) 
Cross-Kerr strength (purple circles) as a function of the detuning $\Delta\,\mathord{=}\,\mathord{-}5$\,MHz. The purple line corresponds to the sum of the yellow and green lines, which exhibit discontinuities at $\Delta\,\mathord{=}\,0$ and $\Delta \,\mathord{\approx}\, \mathord{-} \chi_b$, respectively.}
\label{fig: analytics}
\end{figure}

By moving to a frame rotating at $\omega_a\hat{a}^\dagger \hat{a} + \omega_b\hat{b}^\dagger \hat{b}  + \omega_c\hat{c}^\dagger \hat{c}$ and restricting the coupler basis to the first two energy levels $\{ |g\rangle, |e \rangle \}$, we find that the drive frequency
\begin{equation} 
    \omega_d=\omega_c + \omega_b -\chi_{b} -\omega_a +\Delta,
    \label{Eq: frequency condition}
\end{equation}
where $\Delta$ is the detuning to the resonance condition, successfully actives the dynamics given by 
\begin{equation}
    H_d(t)=g_1 e^{it\Delta} \hat a^\dagger \hat b |g\rangle\langle e| + g_1^* e^{-it\Delta} \hat a \hat b^\dagger \hat |e\rangle\langle g|,
    \label{Eq: Hd}
\end{equation}
which corresponds to Eq.\,2 in the main text. The strength of this interaction is given by 
 \begin{equation}
 g_1=E_J \phi_c^2\phi_a \phi_b \xi^* \approx \sqrt{\chi_{a} \chi_{b}} \xi^*. 
 \label{Eq: g1 derivation}
 \end{equation}
We measured the coherent exchange rate, $g_1$, for different effective drive strengths $|\xi|$ obtained from the coupler AC Stark shift, as shown in Fig.~\ref{fig: analytics}(a). The slope of the interaction rate is in very good agreement with the prediction of Eq.\,\ref{Eq: g1 derivation} using the experimentally measured dispersive shifts (solid line).

To find an analytical equation for the engineered cross-Kerr, we diagonalize the interaction term of Eq.\,\ref{Eq: Hd}, by applying a time-independent Schrieffer-Wolff transformation (SWT)~\cite{schrieffer1966relation,koch2007charge} in the rotating frame of the drive. This is equivalent to applying a time-dependent SWT in the rotating frame of the coupler, which is also equivalent to performing a second-order RWA~\cite{mundhada2017generating}.

In the rotating frame of the drive, the diagonal part of the Hamiltonian reads
\begin{equation}
\begin{split}
\hat H'_0/\hbar &=  \omega_a~\hat{a}^\dagger \hat{a} -\frac{K_a}{2}\hat{a}^\dagger\hat{a}^\dagger\hat{a}\hat{a} \\ 
        & + \omega_b~\hat{b}^\dagger \hat{b} -\frac{K_b} {2}\hat{b}^\dagger\hat{b}^\dagger\hat{b}\hat{b} \\ 
        & + \left(\omega_c-\omega_d\right)~\hat{c}^\dagger \hat{c} -\frac{\alpha}{2}\hat{c}^\dagger\hat{c}^\dagger\hat{c}\hat{c} \\
        & - \chi_{a}~\hat{a}^\dagger\hat{a} ~\hat{c}^\dagger\hat{c} -\chi_{b}~\hat{b}^\dagger\hat{b} ~\hat{c}^\dagger\hat{c} -K_{ab}~\hat{a}^\dagger\hat{a} ~\hat{b}^\dagger\hat{b},
\end{split}
\end{equation}
while the perturbation is given by
\begin{equation}
    \hat V/\hbar=g_1 \hat a^\dagger \hat b |g\rangle\langle e| + g_1^* \hat a \hat b^\dagger \hat |e\rangle\langle g|.
    \label{Eq: V}
\end{equation}

\begin{widetext}
The SWT consists of finding an anti-Hermitian generator $\hat S=-\hat S^\dagger$ that satisfies 
\begin{equation}
    \hat V=\left[ \hat H'_0, \hat S \right],
    \label{Eq: SWT condition}
\end{equation} 
which transforms the Hamiltonian to 
\begin{equation} 
    \hat H'=e^{\hat S} \left(\hat H'_0 + \hat V \right)e^{-\hat S}
   =H'_0 + \hat V + \left[\hat S, \hat H'_0\right] + \left[\hat S,\hat V\right] + \frac{1}{2!}\left[\hat S,\left[\hat S, \hat H'_0\right]\right] + \cdots 
    \approx \hat H'_0 + \frac{1}{2}\left[\hat S, \hat V\right]
\label{Eq: SWT_transformed_H}
\end{equation}

By choosing a generator of the form
\begin{equation}
    \hat S=\beta \left( \hat a^\dagger \hat b |g\rangle \langle e | - \hat a \hat b^\dagger |e\rangle \langle g| \right),
\end{equation}
we can solve Eq.\,\ref{Eq: SWT condition} to find the expression for $\beta$. Since the commutator
\begin{equation*}
    \left[\hat H'_0, \hat a^\dagger \hat b |g\rangle \langle e| \right]=\left( \omega_d -\omega_c + \omega_a - \omega_b -K_a (\hat n_a-1) + K_b \hat n_b+ \chi_{a} (\hat n_a -1) + \chi_{b} (\hat n_b +1)- K_{ab} (\hat n_b - \hat n_a +1) \right) ~\hat a^\dagger \hat b |g\rangle \langle e|
\end{equation*}
depends on the number of photons in both cavities, we expand the transformation generator as 
\begin{equation}
    \hat S=\sum_{n_a,n_b} \beta_{n_an_b} \left( |n_a, n_b, g\rangle \langle n_a-1, n_b+1, e| - |n_a-1, n_b+1, e \rangle \langle n_a, n_b, g| \right),
\end{equation}
where
\begin{equation}
    \beta_{n_a n_b} =\frac{\sqrt{n_a (n_b+1)} ~g_1}{\omega_d -\omega_c + \omega_a - \omega_b -K_a ( n_a-1) + K_b  n_b+ \chi_{a} ( n_a -1) + \chi_{b} ( n_b +1)- K_{ab} ( n_b - n_a +1)}.
\end{equation}
Finally, solving Eq.\,\ref{Eq: SWT_transformed_H} leads to
\begin{equation}
    \hat H'=\hat H'_0 + \sum_{n_a,n_b} \beta_{n_a n_b}\sqrt{n_a (n_b+1)}\, g_1 
    \left( |n_a, n_b, g\rangle\langle n_a, n_b, g|
     - | n_a-1, n_b+1, e \rangle \langle  n_a-1, n_b+1, e|\right).
    \label{Eq: SWT final H}
\end{equation}
\end{widetext}

With these analytical results, we can see that when we parametrically drive this dynamics, the energies of the states $|1\rangle|0\rangle|g\rangle$ and $|1\rangle|1\rangle|g\rangle$ are shifted by 
\begin{align}
    \Delta E_{10} &= \frac{g_1^2}{\Delta},
    \label{Eq: 01 rate}\\
    \Delta E_{11} &= \frac{2\,g_1^2}{\Delta+\chi_b+K_b-K_{ab}} \approx \frac{2\,g_1^2}{\Delta+\chi_b}.
    \label{Eq: 11 rate}
\end{align}
Eq.\,\ref{Eq: 01 rate} shows that in presence of the parametric drive, despite Bob being in vacuum, the state $|\mathord{+}\rangle|0\rangle|g\rangle$ will acquire a phase at a rate $\Delta E_{10}$. Hence, it is important to track the frame of the oscillators (on software) for any subsequent operation. For instance, the frame is tracked as a virtual detuning applied to the displacement pulses used for Wigner tomography. The phase acquired by the state $|\mathord{+}\rangle|0\rangle|g\rangle$ was also experimentally measured as a function of the detuning $\Delta$, see green markers in Fig.\,\ref{fig: analytics}(b), which agrees with the prediction of Eq.\,\ref{Eq: 01 rate} (solid green line). 
Finally, the total cross-Kerr induced in the 0/1 photon subspace is thus

\begin{equation}
\begin{split}
    g_{ab}  &= E_{11}-E_{10}-E_{01}+E_{00} \\ 
            &=  -K_{ab} + \Delta E_{11} - \Delta E_{10} \\
            &= -K_{ab} + \frac{g_1^2}{\Delta}\frac{\Delta-(\chi_{b}+K_b-K_{ab})}{\Delta+(\chi_{b}+K_b-K_{ab})},
\end{split}
\label{Eq: cross_kerr_after_SWT}
\end{equation}
where $E_{ij}$ refers to the energy of the state $|i\rangle|j\rangle|g\rangle$ and $K_{ab}$ is the always-on residual cross-Kerr. We measured the cross-Kerr $g_{ab}$ by tracking the oscillations of $|\mathord{+}\rangle|1\rangle|g\rangle$ in the frame where $|\mathord{+}\rangle|0\rangle |g\rangle$ does not oscillate, at different drive powers for a fixed detuning of $\Delta=-5\,$\,MHz, see blue markers in Fig.\,\ref{fig: analytics}(c). Notice that at different drive powers, the AC Stark shifted frequency of the coupler is different, which subsequently modifies the coherent exchange frequency $\omega_e$. The solid blue line corresponds to the analytical behavior of Eq.\,\ref{Eq: cross_kerr_after_SWT}, which underpredicts the measured interaction strength. We believe this discrepancy arises because the analytical model does not account for the enhanced hybridization between the modes under the strong drive, which renormalizes the dispersive shifts and the detunings. A full Floquet simulation (orange triangles) reproduces the experimental observations in very good agreement. For more details on the Floquet simulation, see Section \ref{Section: Floquet simulations}.

The analytical model still qualitatively captures the asymmetric behavior of the engineered cross-Kerr between positive and negative detunings, see purple markers in Fig.\,\ref{fig: analytics}(d). The green (yellow) dotted line corresponds to $-\Delta E_{10} ~ \left(+\Delta E_{11}\right) $, which diverges at $\Delta=0  ~ \left(\Delta \,\mathord{\approx}\, \mathord{-}\chi_b\right)$. The solid purple line is the sum of both, which qualitatively reproduces the main features observed in the experimental data. For a quantitative agreement, it is desirable to model the system using Floquet theory, as shown in Fig.\,2(b) of the main text.

\section{Floquet simulations}
\label{Section: Floquet simulations}

In the presence of strong parametric drives, the system is governed by a time-periodic Hamiltonian in which the drive-induced shifts of the energy level give rise to unintended multi-photon transitions between the oscillators and higher levels of the coupler, beyond the coherent exchange interaction considered analytically. This scenario is appropriately addressed within the Floquet formalism~\cite{grifoni1998driven, you2025floquet, lu2023high}. We perform Floquet simulations using the QuTiP python package~\cite{lambert2026qutip} to analyze the effects of the parametric drive on the subsystem composed of the SQUID coupled to both cavities, ignoring the side auxiliary transmons. The Hilbert spaces are truncated to dimension $D\,\mathord{=}\,12$ for the SQUID and $D\,\mathord{=}\,4$ for the cavities. The full Hamiltonian considered in the simulation is 
\begin{equation}
\begin{split}
    H/\hbar&= 4 E_C \hat n^2 - E_J \left(\Phi_\text{ext}\right) \cos \hat\phi \\
    & + \omega_a \hat a^\dagger \hat a + \omega_b \hat b^\dagger \hat b \\
    & - g_{ac} (\hat a^\dagger-\hat a) (\hat{c}^\dagger-\hat{c}) \\
    & - g_{bc} (\hat b^\dagger-\hat b) (\hat{c}^\dagger-\hat{c}) \\
    & + \epsilon \cos \left(\omega_d t\,\right)  (\hat{c}^\dagger +\hat{c}),
\end{split}
\end{equation}
where the coupler charging and Josephson energies are those obtained by fitting the experimental data, and summarized in Table\,\ref{tab: SQUID params}; the oscillator frequencies are those measured experimentally through two-tone spectroscopy; the capacitive coupling strength between the oscillators and the SQUID are taken as $g_a/2\pi =43.07$\,MHz and $g_b/2\pi= 22.56$\,MHz, to reproduce the experimentally measured dispersive shifts; and the drive amplitude $\epsilon$ is calibrated to reproduce the measured coupler AC Stark shift. 

The 200\,ns $\cos^2$ ramps for the parametric drive ensure that the undriven eigenstates adiabatically map onto Floquet modes. For a given drive frequency $\omega_d$ and amplitude $\epsilon$, we label each Floquet mode by the eigenstate of the undriven system with which it has the highest overlap. The Floquet quasienergies, defined modulo $\omega_d$, are unfolded into real energies by adding or subtracting integer multiples of $\omega_d$ such that they lie within $\pm 20\%$ of the energy of their corresponding undriven eigenstate. The cross-Kerr between the cavities is then obtained as the difference in the transition energy of Alice when Bob is in vacuum or in $|1\rangle$. The cross-Kerr strength obtained from Floquet simulations, as a function of drive amplitude and detuning, is in very good agreement with the experimental data, see triangle markers in Fig.\,\ref{fig: analytics}(c) and solid line in Fig.\,2 of the main text, respectively. In the latter case, deviations at large detunings arise from the proximity of a single-cavity transition between $|0\rangle|1\rangle|g\rangle$ and $|0\rangle|0\rangle|f\rangle$, which significantly enhances Bob's self-Kerr and its cross-Kerr coupling to Alice~\cite{zhang2022drive}. We attribute this discrepancy to uncertainties in some Hamiltonian parameters, particularly $g_b$, which were calibrated solely from independently measured experimental observables without additional fitting.

\section{Calibration of coherent exchange rate}
\begin{figure}[h!]
\centering
\includegraphics {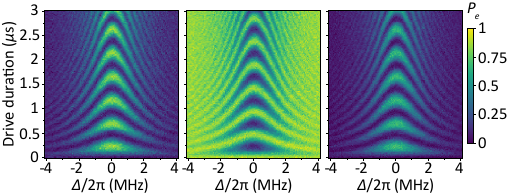}
\caption{\textbf{Coherent exchange of excitations between the coupler and the oscillators}.
Left: probability of finding Alice in vacuum. Middle: probability of finding Bob in vacuum. Right: probability of the coupler being excited. System was initialized in $|1\rangle|0\rangle|g\rangle$. $\Delta\,\mathord{=}\,0$ corresponds to $\omega_d\,\mathord{=}\,\omega_e$, where exchange occurs in $\sim$\,200\,ns. The coupler configuration corresponds to Fig.\,2(a) in the main text.}
\label{fig:chevron}
\end{figure}

To confirm the location of the coherent exchange interaction $\omega_e$, and calibrate the exchange strength, $g_1$, we measure the population of Alice, Bob, and the coupler, as a function of the drive duration and frequency ($\omega_d$), as shown in Fig.\,\ref{fig:chevron}.

The experiment starts by preparing the state $|1\rangle|0\rangle|g\rangle$ through a numerical GRAPE pulse on the side transmon $q_a$ and Alice. A microwave drive is applied to the coupler with variable duration $t$ and frequency $\omega_d$. After the drive pulse, the system populations are measured. The coupler population is obtained via direct dispersive readout of its readout resonator, while the cavity populations are measured using 
selective $\pi$-pulses on their auxiliary transmon, conditioned on each cavity being in vacuum. 
The cavity data is normalized by the selective $\pi$-pulse contrast, and the coupler population is corrected by the readout contrast.

The data exhibits the characteristic chevron pattern expected for a coherent exchange between $|1\rangle|0\rangle|g\rangle$ and $|0\rangle|1\rangle|e\rangle$. At zero detuning, the populations oscillate sinusoidally in time with a frequency $2g_1$ while detuning from the exchange condition leads to faster oscillations with reduced contrast. A line cut yields the oscillations shown in Fig.\,2(a) of the main text, from which we can extract $g_1/2\pi\,\mathord{=}\,1.024 \pm 0.004$\,MHz.

\section{Extension to multi-photon codes}
\begin{figure}[tbh!]
\centering
\includegraphics{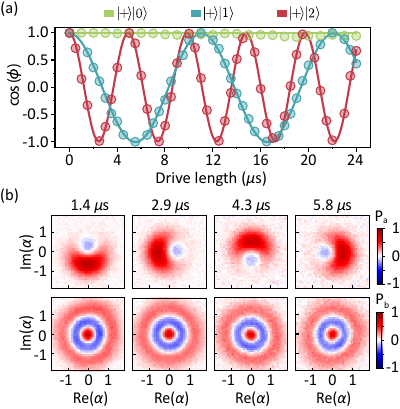}
\caption{\textbf{Bosonic enhancement of the gate rate.} (a) Phase acquired by the target cavity initialized in $|\mathord{+}\rangle$ during the gate when the control cavity is prepared in $|0\rangle$, $|1\rangle$ and $|2\rangle$. The gate time is reduced from $\sim$\,5.5$\mu$s to $\sim$\,2.5\,$\mu$s when populating the control cavity with two photons. (b) Representative Wigner tomography snapshots of the target cavity at selected times for the initial state $|\mathord{+}\rangle|2\rangle$.}
\label{fig:2_plus_wigner}
\end{figure}

Due to the bosonic enhancement factor in the coupling strength, we can increase the gate speed by populating higher Fock states in the control cavity. We demonstrate this by measuring the phase acquired by the $|\mathord{+}\rangle$ state during gate operation when the control cavity is in $|2\rangle$ and comparing it to the equivalent case when the control is in $|1\rangle$. The phases acquired as a function of time for both cases are shown in Fig.\,\ref{fig:2_plus_wigner}(a). Representative Wigner tomography snapshots at selected times for initial state $|\mathord{+}\rangle|2\rangle$ are shown in Fig.\,\ref{fig:2_plus_wigner}(b). We observe that the CZ gate time is reduced from $\sim$\,5.5\,$\mu$s to $\sim$\,2.5\,$\mu$s. 

It is important to note that the gate time is reduced by more than a factor of two as the resonant condition is shifted by $n\chi_b$ when the control cavity is prepared in $|n\rangle$ in accordance with Eq.~\ref{Eq: frequency condition}. Since the gate detuning is defined with respect to the single-photon case and is kept constant for both experiments, the effective detuning becomes smaller for the two-photon state, leading to a gate rate that exceeds the simple factor-of-two scaling expected from bosonic enhancement alone. This entails that when the engineered dynamics is applied to multi-photon states, we must account for this difference in effective detuning appropriately. 

\section{Pauli bars for other states}
\begin{figure*}[tbh!]
    \centering
    \includegraphics{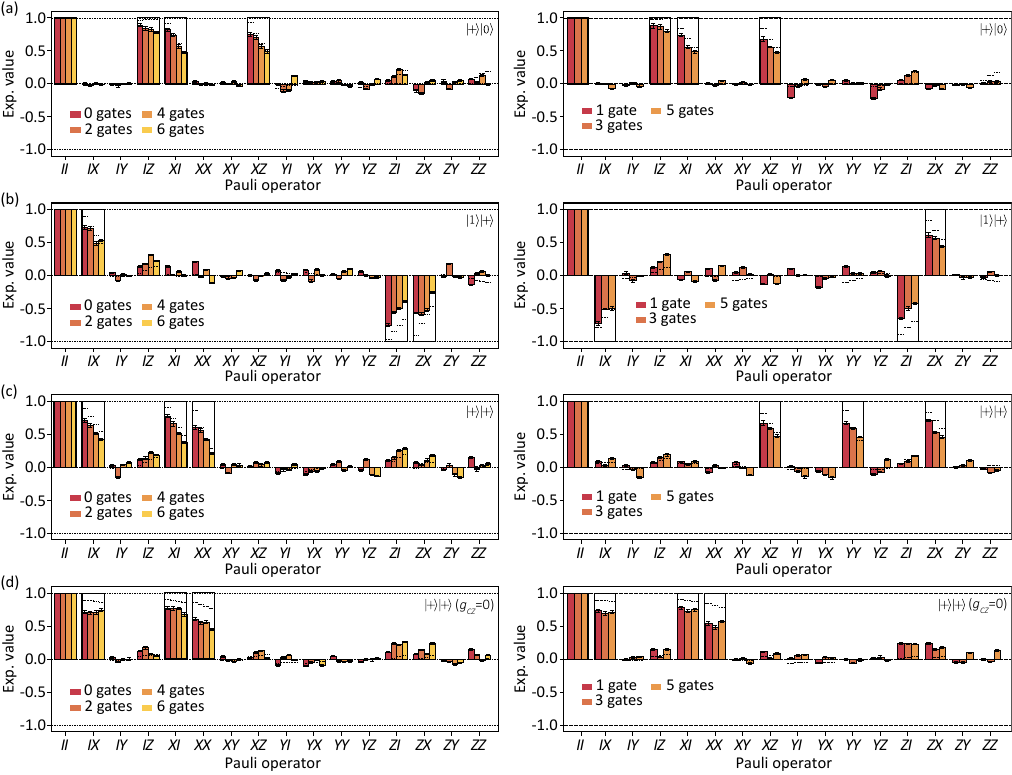}
    \caption{\textbf{Reconstructed density matrices after several CZ gates.} Reconstructed two-cavity states in the Pauli operator representation after 0–6 applied CZ gates, for the initial states (a) $|\mathord{+}\rangle|0\rangle$, (b) $|1\rangle|\mathord{+}\rangle$ (b), and (c) and (d) $|\mathord{+}\rangle|\mathord{+}\rangle$ with drive on and off, respectively. Black outlines indicate the ideal values and dashed lines the numerical simulations including cavity decoherence.}
    \label{fig:additional pauli bars}
\end{figure*}

Here, we present the reconstructed two-cavity density matrices in the Pauli decomposition, after applying 0 to 6 CZ gates to the initial states $|\mathord{+}\rangle |0\rangle$, $|1\rangle|\mathord{+}\rangle$ and $|\mathord{+}\rangle|\mathord{+}\rangle$, see Fig.\,\ref{fig:additional pauli bars}. For the initial state $|\mathord{+}\rangle|\mathord{+}\rangle$ we further distinguish between the case of drive on and off (i.e. $g_{CZ}\,\mathord{=}\,0$). These data complement Fig.\,3 of the main text, where the state fidelity is shown as a function of applied gates.

The density matrices are reconstructed from the measurement data using ORENS and Bayesian inference, see Section \ref{Section: Quantum state reconstruction} for further details. The colored bars correspond to the experimentally reconstructed Pauli coefficients. The black outlined bars indicate the ideal values, while the dashed lines show the expected coefficients obtained from numerical simulations. The simulations are performed using the effective Hamiltonian $H_{\mathrm{eff}}\,\mathord{=}\,g_{ab}\, a^{\dagger} a \,b^{\dagger} b$ with a gate time of $5.5\,\mu$s. We perform the simulations including the experimentally-calibrated cavity decoherence parameters when the drive is on, with $T_{1a}\,\mathord{=}\,210\,\mu$s, $T_{\phi a}\,\mathord{=}\,70\,\mu$s, $T_{1b}\,\mathord{=}\,388\,\mu$s and $T_{\phi b}\,\mathord{=}\,52\,\mu$s. See Section \ref{Section: driven decoherence} for further details.

\begin{figure}[tbh!]
    \includegraphics{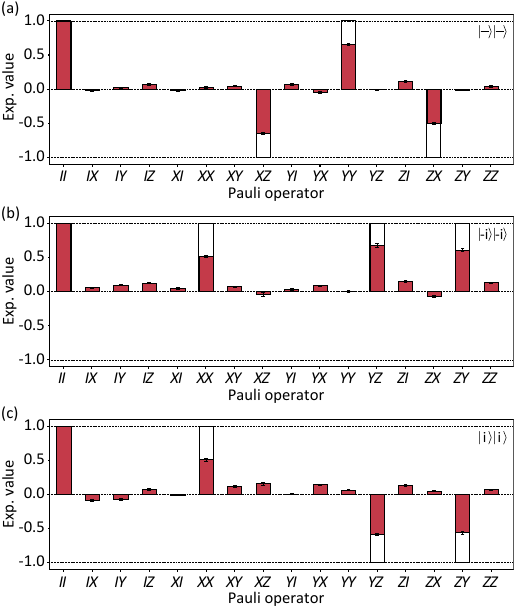}
    \caption{\textbf{Reconstructed density matrices for other cardinal states.} Reconstructed two-cavity states in the Pauli operator representation after a single CZ gate applied to the initial states (a) $|\mathord{-}\rangle|\mathord{-}\rangle$, (b) $|\mathord{-}i\rangle|\mathord{-}i\rangle$, and (c) $|i\rangle|i\rangle$. Black outlines correspond to ideal states, and dashed lines to numerical simulations including cavity decoherence.}
    \label{fig:additional pauli bars cardinal}
\end{figure}

Furthermore, Fig.\,\ref{fig:additional pauli bars cardinal} shows the reconstructed Pauli decompositions after a single CZ gate applied to the initial states $|\mathord{-}\rangle|\mathord{-}\rangle$, $|i\rangle|i\rangle$ and $|\mathord{-}i\rangle|\mathord{-}i\rangle$, which ideally produce maximally entangled states under the engineered interaction. 
As above, the Pauli coefficients are obtained from measurements using ORENS combined with Bayesian inference. The experimental results are compared to both the ideal states (black outline) and to numerical simulations including decoherence (dashed lines) using the same effective Hamiltonian and parameters described above.

\section{Quantum state reconstruction}
\label{Section: Quantum state reconstruction}
We perform quantum state reconstruction following the Optimized Reconstruction with Excitation Number Sampling (ORENS) protocol. We extended the scheme in Ref.~\cite{krisnanda2025demonstrating} for reconstructing two-mode cavity states. It involves applying a set of optimized displacement points on both cavities and performing photon number measurements. The measurement results are then processed to estimate the two-mode density matrix. Up to this point, the resulting density matrix might not be physical due to the presence of noise in the experimental data. Thus, we employ Bayesian inference to get a statistically accurate and physical estimated density matrix. The following two sections provide details on the implementation of the ORENS protocol and the Bayesian inference procedure used to obtain the final estimated density matrix.

\subsection{State reconstruction via ORENS}
ORENS enables efficient quantum state tomography using the theoretical minimum number of measured single-valued observables, which are excitation (photon) number measurements performed after a set of optimized phase-space displacements. For reconstructing a single bosonic mode of dimension $D$, the least number of observables is $N_{\text{obs}}\,\mathord{=}\,D^2-1$. These are obtained using $N_{\text{obs}}$ optimized displacement points, each followed by a measurement of a single photon number $n$. The set of observables can be written as $X_i\,\mathord{=}\,\text{tr}(|n\rangle \langle n|\hat D^{\dagger}(\alpha_i)\rho \hat D(\alpha_i))$, where $i\,\mathord{=}\,1,2,\cdots,N_{\text{obs}}$ and $\{\alpha_i\}$ is the set of optimized displacement points obtained by minimizing the measurement matrix via a gradient descent algorithm~\cite{krisnanda2025demonstrating}.

An arbitrary density matrix $\rho$ of dimension $D$ can be parametrized by its $D^2$ real parameters (without the normalization condition). In our case, we take the diagonal elements of $\rho$ as well as its real and imaginary off-diagonal (upper triangular) elements, which we shall arrange in a vector form $\vec Y$. The parameters ($\vec Y$) and observables (also arranged in a vector form $\vec X$) are linearly related~\cite{krisnanda2023tomographic}. In our case, we write this as $\vec Y\,\mathord{=}\,M\vec X$. Here, $M$ is a mapping matrix, which can be computed given the set of optimized displacement points $\{\alpha_i \}$ and photon number $n$ to be measured. To get an estimate of the parameters $\vec Y_{\text{est}}$, given the experimental data $\vec X$, we apply the inverse of the matrix equation $\vec Y_{\text{est}}\,\mathord{=}\,M^+ \vec X$ with $M^+\,\mathord{=}\,(M^{\dagger}M)^{-1}M^{\dagger}$ being the left Moore-Penrose pseudoinverse. From the estimated parameters, we then construct the density matrix $\rho_{\text{LS}}$. Given this density matrix, which might be unphysical, we employ Bayesian inference to get the final physical density matrix (see next section).

To reconstruct a two-mode cavity state in the present case, we use a total dimension of $D\,\mathord{=}\,4$ (2 for each mode) and $D\,\mathord{=}\,9$ (3 for each mode) for the case of 0/1 encoding and 0/2 encoding, respectively. This truncation is justified by the absence of any observed population in $|2\rangle$ and $|3\rangle$ after the gate, within readout error. The displacement operations we apply are of the form $\{\hat D_a(\alpha_i)\otimes \hat D_b(\alpha_j)\}$, where $\{\alpha_i\}$ refers to the optimized displacements obtained via optimization for the single-mode tomography case. For the 0/1 and 0/2 encoding, we use a total of $25$ and $100$ different displacements, respectively. This exceeds the theoretical minimum of $D^2-1$ in both encoding scenarios, providing an overcomplete set of observables that improves robustness to uncertainties and reduces the averaging required per observable. After each displacement, we measure the probability of observing the joint photon number $|n\rangle_a\otimes |n\rangle_b$ in the two-mode cavity state. This is obtained through individual single-shot measurements of photon number $|n\rangle$ in Alice and Bob, by applying selective $\pi$-pulses on their side transmons at frequencies $\omega_{q_a}\mathord{-}n\chi_{a\,q_a}$ and $\omega_{q_b}\mathord{-}n\chi_{b\,q_b}$, respectively, and measuring the resulting excited state probabilities. By correlating the outcomes of these measurements for each experimental run, we obtain the joint probability of observing $|n\rangle_a\otimes |n\rangle_b$ photons. 

Note that as we use parametrization with $D^2$ parameters (without enforcing the normalization condition), the density matrix obtained via linear inversion (least square) $\rho_{\text{LS}}$ might not have unit trace and positive eigenvalues. We shall keep $\rho_{\text{LS}}$ as a data-driven estimator that faithfully reflects the measurement outcomes, and use it solely as the reference point for a subsequent Bayesian inference procedure that restores physicality in the final estimated density matrix. 

Two-mode state reconstruction has previously also been achieved by measuring the joint parity $\langle \hat P_a\otimes \hat P_b \rangle$~\cite{wang2016schrodinger}. However, this approach requires closely matched dispersive shifts $\chi$ between transmon and both cavities, placing stringent constraints on the hardware. One can in principle perform separated single shot parity measurements for $\hat P_a$ and $\hat P_b$, and correlate them to get $\langle \hat P_a\otimes \hat P_b \rangle$, as we did for photon number in this work. However, the traditional parity measurement $\pi/2$-wait-$\pi/2$ is known to suffer from coherent as well as incoherent errors~\cite{krisnanda2025demonstrating}. A standard technique employed to counter this problem is to perform a corrected parity measurement, in which $\langle \hat P\rangle_{\text{corr}}\,\mathord{=}\,(\langle \hat P\rangle - \langle \hat P\rangle_{\text{rev}})/2$ is measured. Here $\langle \hat P\rangle_{\text{rev}}$ is obtained using a $\pi/2$-wait-$(-\pi/2)$ sequence. Implementing this technique in our device would require correlating 4 different measurements: $\hat P_a$ with $\hat P_b$; $\hat P_a$ with $\hat P_{B,\text{rev}}$; $\hat P_{A,\text{rev}}$ with $\hat P_b$; and $\hat P_{A,\text{rev}}$ with $\hat P_{B,\text{rev}}$, therefore multiplying the number of required measurements by a factor of $4$.
This tomography method adds a factor of $2^N$ to the total number of displacement points for $N$-mode state reconstruction, which is certainly less favorable compared to the method implemented in this work.

\subsection{Bayesian inference}

To obtain a physical (positive semidefinite) estimated density matrix, a standard technique consists of performing linear inversion (as explained in the previous section) followed by maximum likelihood estimation~\cite{smolin2012efficient}. However, this method suffers from the problem, that the estimated physical density matrix often has zero eigenvalues~\cite{blume2010optimal}. This is not justified by finite measurement statistics, as the absence of observed events cannot support the conclusion that the corresponding probabilities are exactly zero. Instead, we employ Bayesian inference, which gives a statistically more accurate estimate. Incorporating Bayes' rule, Bayesian inference provides a posterior probability distribution given the measured experimental data and prior knowledge. The posterior distribution can then be used (sampled) to get an estimate for any function of the density matrix. We follow an efficient Bayesian inference protocol introduced in Ref.~\cite{lukens2020practical}.

We use uniform prior, i.e., no prior assumption on the density matrix, such that the posterior distribution is proportional to the pseudo likelihood function
\begin{equation}
\mathcal{L}(\rho) \propto \exp\Biggr(-\frac{N}{2}||\rho - \rho_{LS}||^2\Biggr)
\end{equation}
where $||\cdot||$ is the Frobenius norm and $N$ is the total number of events (number of repetitions times number of applied displacements). We then draw a total of $R\,T$ samples according to the Crank-Nicolson Metropolis-Hastings procedure, with a thinning parameter $T=2^7$ to reduce serial correlation in the subsequent samplings and keep only $R=2^{10}$ samples. The final reconstructed state is given by the Bayesian mean estimator: 
\begin{equation}
\rho_{\text{BME}}=\frac{1}{R} \sum_{r=1}^{R}\rho_r
\end{equation}
where $\{\rho_r\}$ are the retained physical density matrices sampled from the posterior distribution. We calculate the fidelity between the density matrix $\rho_{\text{BME}}$ and a target density matrix $\rho_{tar}$ using: 
\begin{equation}    \mathcal{F}=\Biggr(Tr\sqrt{\sqrt{\rho_{\text{tar}}}\rho_{\text{BME}}\sqrt{\rho_{\text{tar}}}}\Biggr)^2,
\end{equation}
where $\rho_{\text{tar}}$ is the target density matrix generated by simulating the numerical GRAPE pulses under realistic decoherence mechanisms, followed by propagation through an ideal gate model with no decoherence, where the effective Hamiltonian used is $\hat H_{\text{eff}}\,\mathord{=}\,g_{ab}\,\hat a^{\dagger}\hat a \,\hat b^{\dagger}\hat b$. This procedure isolates the sources of gate imperfections, such as system decoherence and spurious coupler excitations, from errors of the state preparation and measurement. A quantitative error budget is provided in Section~\ref{section: errors}.

\section{Drive-induced decoherence}
\label{Section: driven decoherence}
\subsection{Strong drive effects on cavity coherences}
In this work, we drive the coupler with a strong microwave tone to activate the cross-Kerr interaction. When operating in the Raman regime, the coupler is never physically populated, thereby mitigating errors arising directly from its decoherence mechanisms. Nevertheless, we observe a substantial reduction of the oscillator coherences as a function of both the drive strength and the detuning from the resonance condition described in the main text. To quantify the cavity properties under this strong drive, we characterize the driven single-photon lifetime of the oscillators by loading a photon in Alice (Bob) and replacing the delay time by a strong off-resonant drive applied to the coupler. We observe a degradation of $T_1$ times from 700–900$\,\mu$s (800–1000$\,\mu$s) to $\sim$\,210\,$\mu$s ($\sim$\,390\,$\mu$s). The dephasing time of the oscillators was characterized via a Ramsey experiment where Alice (Bob) is initialized in a superposition $(|0\rangle + |1\rangle)/\sqrt{2}$ and the delay time is replaced with a strong off-resonant drive. We observe a reduction in the oscillator $T_2^*$ times from 500–600$\,\mu$s (700–900$\,\mu$s) to $\sim$\,40\,$\mu$s ($\sim$\,60\,$\mu$s). 

The observed reduction in coherence is consistent with previous reports in strongly driven cQED systems~\cite{carde2025flux, kim2025ultracoherent, lu2023high}. We attribute the enhanced oscillator decay to drive-induced dissipation~\cite{you2025floquet, kim2025ultracoherent}, arising from the dressing of the coupler by the off-resonant coherent exchange interaction. A simple model of this effect can be obtained by considering the first-order perturbative correction to the coupler decay operator, $|g\rangle\langle e| \,\mathord{\rightarrow}\, |g\rangle\langle e| + \left[\hat S, |g\rangle\langle e| \right]$, where $\hat S$ is the SWT generator introduced in Section~\ref{Section: SWT}. The resulting commutator generates an additional term $\beta \,a b^\dagger \left(|g\rangle\langle g|-|e\rangle\langle e|\right)$. Likewise, dressing the coupler dephasing operator yields $\beta \left(a^\dagger b |g\rangle\langle e|\mathord{+} a b^\dagger|e\rangle\langle g|\right)$. Operating the coupler away from its sweet spot, where $T^c_\phi\,\mathord{\sim}$\,2\,$\mu$s, may therefore lead to an effective oscillator dressed decay time of $T_1^{a,b}\mathord{\sim}\beta^{-2} T_\phi^{c}\,\mathord{\sim}$\,200\,$\mu$s, consistent with the observed driven coherence times. In addition, the applied drive induces AC Stark shifts of the coupler transitions, which can approach the cavity frequencies at large drive amplitudes. Furthermore, the off-resonant drive enhances hybridization between the modes, thereby increasing the inherited cavity decay~\cite{zhang2019engineering, zhang2022drive}. Similar to~\cite{kim2025ultracoherent}, there is a tradeoff between the gate time, which scales with $\Delta$, and the sideband-dressed oscillator decay, which scales as $\Delta^{-2}$. While this simple model provides a qualitatively consistent description of our observations, a more rigorous treatment is required to systematically account such drive-induced decoherence effects. This remains an active area of research and lies beyond the scope of the present work. 

We further attribute the degradation of the oscillator dephasing time to two distinct mechanisms. First, dressed dephasing~\cite{zhang2019engineering}, arising from an elevated ancilla temperature in the Floquet basis. This contribution depends primarily on the drive-coupler detuning ($\mathord{\sim}$\,280\,MHz) and is therefore insensitive to the interaction detuning ($\Delta\,\mathord{\sim}$\,6\,MHz), resulting in a nearly flat photon-shot noise spectrum~\cite{kim2025ultracoherent}. Second, the interplay between the enhanced oscillator decay rates and the direct cross-Kerr coupling between the oscillators leads to an additional, cross-Kerr dependent dephasing channel.

\subsection{Driven coupler decoherence}
\begin{figure}[t!]
\centering
\includegraphics {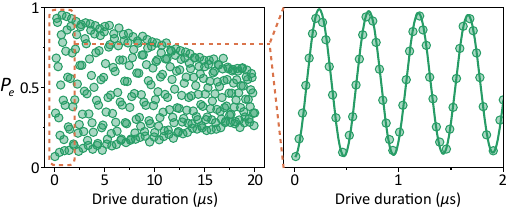}
\caption{\textbf{Coherent coupler oscillations.} Experimentally measured exchange oscillations of the coupler mode (green markers), used to fit Eq.~\ref{Eq: coherent exchange fit} and extract the corresponding decay and dephasing times. The right panel shows a zoomed-in view of the first $2\,\mu$s of the oscillations. The solid line corresponds to master-equation simulations using the cavity and coupler decoherence rates, with the coupler dephasing time measured via a CPMG sequence.}
\label{fig: coupler oscillations}
\end{figure}

When the system is driven at the resonance condition, we observe a coherent exchange of excitations between the states $\{|1\rangle|0\rangle|g\rangle, |0\rangle|1\rangle|e\rangle\}$, as shown in Fig.\,2(a) of the main text. The corresponding coupler population oscillations are presented in Fig.\,\ref{fig: coupler oscillations}. In the presence of decoherence, dephasing reduces the oscillation contrast, while energy relaxation skews the dynamics toward the ground state. Following~\cite{lu2023high}, we model the coupler oscillations as a function of the drive duration using
\begin{equation}
    P_0(t)=Ae^{-\kappa_1t}\Big(1+ e^{-\kappa_{\phi}t}\cos(2g_1t+B)\Big)+C,
    \label{Eq: coherent exchange fit}
\end{equation}
where $A$, $B$, and $C$ are amplitude, initial phase and offset fitting parameters to compensate for state preparation and measurement (SPAM) imperfections; $\kappa_{\phi}$ denotes the effective dephasing rate within the exchange subspace; and $\kappa_1$ characterizes population decay to the vacuum state. From this fit, we extract effective coherence times of $\kappa_1^{-1}\,\mathord{=}\,94 \,\mathord{\pm}\, 3 \,\mu \text{s}$ and $\kappa_\phi^{-1}\,\mathord{=}\,28.0\, \mathord{\pm}\, 0.4 \,\mu \text{s}$. Because the excitations involved in the interaction are delocalized across both cavities and the nonlinear coupler, these rates cannot be directly interpreted as averages of the individual cavity coherence times, in contrast to Ref.~\cite{lu2023high}.

Instead, we perform full master equation simulations using the independently measured cavity and coupler decoherence rates. Importantly, using the bare coupler dephasing time extracted from Ramsey experiments predicts a significantly faster decay of the oscillations than is observed experimentally. This discrepancy indicates that the relevant coherence properties during the exchange process differ from those inferred from static measurements. It is well established that, under continuous driving, the effective coherence times of superconducting transmons are modified relative to their free-evolution values~\cite{fei2013rotating, ithier2005decoherence}. To account for this, we adopt a coupler $T_2$ time of $13\,\mu$s, corresponding to the experimentally measured decoherence time obtained via a Carr–Purcell–Meiboom–Gill (CPMG) sequence~\cite{carr1954effects}. This choice is justified by the fact that the continuous drive effectively rotates the transmon Bloch vector, producing dynamics analogous to a repeated echo sequence. With this modification, the simulations are in very good agreement with the experimental data, as shown in
the right panel of Fig.\,\ref{fig: coupler oscillations} and in Fig.\,2(a) of the main text.

\section{Error budgeting}
\label{section: errors}
In this section, we break down and estimate the different sources of infidelity affecting our experimental data. To this end, we perform full master equation simulations in which individual error mechanisms are introduced one at a time, allowing us to isolate their respective contributions. While this approach provides a useful estimate of each contribution to the infidelity, it does not capture potential correlations between error channels. We therefore also present simulations that incorporate all error sources simultaneously for comparison. Overall, both approaches yield a consistent reproduction of the experimentally measured values.

\begin{figure}[thb!]
\centering
\includegraphics {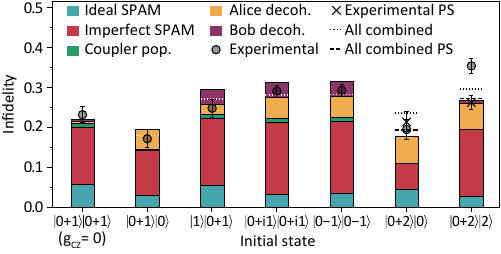}
\caption{\textbf{Error budgeting for the CZ gate.}
Stacked bars show simulated contributions to the infidelity arising from SPAM (with and without decoherence), coupler population, and cavity decoherence for several initial states. Dotted lines indicate simulations that include all error channels simultaneously, while gray markers denote the experimentally measured infidelities. For the last two states, we show the simulated (dashed line) and experimental (crosses) infidelities after post-selecting (PS) the data on the cavities being in an even-parity state.}
\label{fig:Error budget}
\end{figure}

We begin by considering errors arising from SPAM in an idealized setting without decoherence. We simulate the numerical GRAPE pulses used to prepare the initial states using the Hamiltonian of Eq.\,\ref{Eq: side Hams}, followed by an application of a single CZ gate, modeled by the effective Hamiltonian 
\begin{equation}
H_{eff}\, \mathord{=} \,g_{ab} a^{\dagger} a b^{\dagger} b - \chi_a \,a^{\dagger} a \,|e\rangle\langle e| - \chi_b \, b^{\dagger} b \, |e\rangle\langle e|.
\end{equation}
We then simulate the ORENS tomography using a pulse-based description, with realistic selective $\pi$-pulses and displacement operations, while neglecting readout errors. The infidelity of the reconstructed simulated final state is quantified as $1-|\langle\psi_{\text{sim}}|\psi_{\text{target}}\rangle|^2$, where $|\psi_{\text{target}}\rangle$ denotes the ideal state after the gate (e.g. $|1\rangle|\mathord{-}\rangle$ for an initial state $|1\rangle|\mathord{+}\rangle$). The infidelity obtained from this decoherence-free simulation (see turquoise bars in Fig.\,\ref{fig:Error budget}) arises primarily from imperfections in the state preparation, that generate small populations outside the 0/1 subspace, which cannot be captured by the $D\,\mathord{=}\,2$ ORENS tomography. These leakage populations are sufficiently small not to be detected in experiments.

We then repeat the previous analysis, now including decoherence during both the initial state preparation and the ORENS tomography protocol. In addition, measurement errors are incorporated by applying the confusion matrices described in Section~\ref{Section: readout} to the expected excited state populations of the side transmons. This contribution constitutes the dominant source of infidelity in our experiments (see red bars in Fig.~\ref{fig:Error budget}) and is ultimately limited by the coherence times of the side transmons. We also consider infidelities arising from residual coupler population after the gate (green), corresponding to the residual excitations shown as green crosses in Fig.\,2(b) of the main text, as well as from the driven decoherence of Alice (orange) and Bob (purple) during the gate, using the values described in Section~\ref{Section: driven decoherence}. Each of these contributions is simulated by including only a single error mechanism at a time, without accounting for SPAM errors. Finally, we compute the total infidelity by including all decoherence channels simultaneously (dotted lines). In some cases, this results in a lower infidelity than that obtained by summing the individual contributions, indicating that certain error mechanisms are not independent. For instance, cavity decoherence can partially suppress spurious populations outside the tomography subspace, thereby mitigating their contribution to the total infidelity. For the states in the 0/2 subspace, we further simulate the effect of post-selection by projecting the simulated states into the even-parity subspace (dotted lines) before passing them into the ORENS tomography simulation. Overall, for a single gate, the SPAM errors dominate the infidelity, whereas for increasing numbers of gates (Fig.\,3(e) in the main text), cavity decoherence becomes the leading error mechanism. 

\section{Cross-Kerr in 3-wave mixing}

\begin{figure}[!tbh]
\centering
\includegraphics {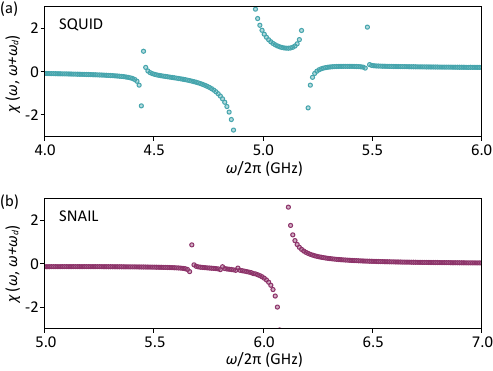}
\caption{
\textbf{SQUID and SNAIL susceptibilities.} 
Real part of the susceptibility of a (a) SQUID, (b) SNAIL, quantifying the coupler response to a probe at $\omega$ when driven at $\omega_d$. The SNAIL, biased at its Kerr-free point, exhibits fewer spurious resonances than the SQUID for the same drive amplitude.
}
\label{fig:SNAIL}
\end{figure}

The Raman-assisted protocol presented in this work employs a SQUID coupler and concatenates two 4-wave mixing processes to engineer an effective 6th-order Hamiltonian term (involving 2 drive photons). While this activates the desired cross-Kerr interaction, it also introduces parasitic effects. The always-on dispersive coupling between the cavities and the coupler modifies the effective drive detuning from the coherent exchange condition as a function of the cavities photon number. In addition, AC Stark shifts arising from the 4-wave mixing dynamics can lead to undesired frequency collisions under strong drives. Finally, the nonlinear mixing between all the modes that participate in the coupler junction gives rise to a spectrum of spurious transitions that constrains the choice of drive parameters. Although spurious transitions involving cavity modes are spectrally narrow, sufficient detuning is required to suppress off-resonant second-order processes.

To mitigate these imperfections, we envision replacing the SQUID coupler with a tailored 3-wave mixing element. For example, a linear inductive coupler (LINC)~\cite{maiti2025linear} or a Superconducting Nonlinear Asymmetric Inductive eLement (SNAIL) biased to the operating point where it only exhibits 3- and 5-wave mixing~\cite{frattini20173}. For the remainder of this section we focus on the latter as a concrete example. The SNAIL Hamiltonian is
\begin{equation}
    H^\text{S}/\hbar\,\mathord{=}\,\omega_s \hat s^\dagger \hat s + g_3 \left( \hat s + \hat s^\dagger \right)^3 + g_5 \left( \hat s + \hat s^\dagger \right)^5 + \cdots,
\end{equation}
where $\hat s$ is the annihilation operator of the SNAIL mode, and $g_3$ and $g_5$ are its third- and fifth-order nonlinearities. Coupling the SNAIL to both oscillators and driving at a frequency $\omega^{(S)}_d = \omega_c - \omega_a$ engineers the interaction
\begin{equation}
    \hat H^{(3)}_d/\hbar = g_3 \left(\hat a^\dagger \hat s  + \hat a \hat s^\dagger\right).
\end{equation}
At this same resonant condition, the 5-wave mixing term
\begin{equation}
    \hat H^{(5)}_d/\hbar = g_5 \left(\hat a^\dagger \hat b^\dagger \hat b\hat s  + \hat a \hat b^\dagger \hat b \hat s^\dagger \right)
\end{equation}
is also resonant. As in the SQUID case, driving with a detuning $\Delta$ from this resonance activates a Raman-assisted interaction between $\hat H^{(3)}_d$ and $\hat H^{(5)}_d$, yielding an effective Hamiltonian 
\begin{equation}
    \hat H^\text{S}_\text{eff}/\hbar =\frac{ g_3 g_5 }{\Delta} \hat a^\dagger \hat a \, \hat b^\dagger \hat b \, \left(|g\rangle\langle g|-|e\rangle\langle e|\right).
\end{equation}
Here we assume that, despite operating at its Kerr-free point, the SNAIL retains a small but finite anharmonicity from its 3-wave mixing term. Provided the drive amplitude and bandwidth remain small compared to this anharmonicity, leakage to higher excited states is suppressed, and the coupler can be approximated as an effective two-level system, with $\hat{s} = |g\rangle\langle e|$.

Using a 3-wave mixing element would not only solve the problem of photon-number–dependent cross-Kerr couplings, but also significantly reduce drive-induced frequency shifts and suppress many of the spurious multi-photon processes inherent to 4-wave mixing. To illustrate the latter, we compare the susceptibility $\chi(\omega, \,\omega\mathord{+}\omega_d)$ of a SQUID and a SNAIL coupler, following the procedure of~\cite{zhang2022drive}, see Fig.~\ref{fig:SNAIL}. Here, the susceptibility characterizes the frequency-converting response of the coupler, relating a probe at frequency $\omega$ to a response at $\omega\mathord{+}\omega_d$, where $\omega_d$ is the drive frequency. When the cross-Kerr drive is on, the SNAIL exhibits fewer activated resonances across the probe frequency spectrum, resulting in a significantly cleaner susceptibility profile.

\bibliography{references_combined_v2}

\end{document}